\documentclass{elsart}
\usepackage{epsfig}
\usepackage{graphicx}
\usepackage{lscape}
\usepackage{subfigure}
\usepackage{amsmath}
\usepackage{amssymb}
\usepackage{color}
\journal{Journal of Magnetism and Magnetic Materials}
\begin{document}
\begin{frontmatter}
\title{Critical Behavior of AC Antiferromagnetic and Ferromagnetic Susceptibilities of a Spin-$1/2$ Metamagnetic Ising System }
\author{Gul Gulpinar  $^{ a}$ },
%corauthref{cor}
\corauth[cor]{corresponding author.}
\ead{gul.gulpinar@deu.edu.tr}
\author{Erol Vatansever $^{a}$}
\address{ Dokuz Eylul University, Department of Physics, 35160-Buca, Izmir, Turkey}
%**************************************************************************************************************************
%**************************************************************************************************************************
\begin{abstract}
\vskip0.2cm \hskip1.1cm
 In this study, the temperature variations of  the  equilibrium and the non-equilibrium  antiferromagnetic  and ferromagnetic susceptibilities of a metamagnetic system are examined near the critical point.  The kinetic equations describing the time dependencies of the total and staggered magnetizations are derived by utilizing linear response theory. In order to obtain dynamic magnetic relaxation behavior of the system,  the stationary solutions of the kinetic equations in existence of  sinusoidal  staggered and physical external magnetic fields are performed. In addition, the static and dynamical mean field critical exponents are calculated in order to formulate the critical behavior of antiferromagnetic and ferromagnetic magnetic response of a metamagnetic system. Finally, a comparison of the findings of this study  with previous theoretical and experimental studies is represented and it is shown that a good agreement is found with our results.
\end{abstract}
%**************************************************************************************************************************
%**************************************************************************************************************************
\end{frontmatter}
%**************************************************************************************************************************
%**************************************************************************************************************************
\section{Introduction}
\vskip0.2cm \hskip1.1cm

Investigation of the AC or dynamic susceptibilities are the most preferred techniques
to study Magnetic relaxation (MR) which is related to magnetic
hysteresis and present in all stages of the development of
ordered phase is an important approach to probe magnetic systems. They are obtained from the dynamic response of the system to time-dependent magnetic field
and now commonly used to investigate the magnetic properties of
high-$T_{c}$ systems \cite{Engelstad}, cobalt-based alloys \cite{Durin}
, nanoparticles \cite{Raap}, spin glasses \cite{Kötzler}, molecule-based magnets  \cite{Girtu},
and magnetic fluids \cite {Fannin}.
Theoretical investigation of dynamic magnetic response of the Ising models
has been an subject of interest for quite a long time: In 1966,
Barry has studied spin-$1/2$ Ising model
by a method combining statistical theory of phase transitions and
irreversible thermodynamics \cite{Barry66}. Using the same method,
Barry and Harrington has focused on the theory of relaxation phenomena in an antiferromagnet \cite{Barry71}.
In addition, they obtained the temperature and frequency dependencies of the
magnetic absorption and dispersion factor in the neighborhood  of the critical temperature.
On the other hand, Suzuki and Kubo has obtained the time dependent susceptibility of the
kinetic Ising model \cite{Suzuki}. Acharyya and Chakrabarti
presented the real and imaginary parts of magnetic
susceptibility near the order–disorder transition point
of a spin-$1/2$ Ising system in the presence of
a periodically varying external field by using Monte-Carlo
simulations \cite{Acharyya}.
On the other hand, The dynamic magnetic response of the materials and the development of methods for its
modification are important for their potential applications: Cores made of cobalt-based alloys in low signal detectors of gravitational physics contribute as a noise source with a spectral density proportional to the ac susceptibility
of the alloy \cite{Durin}; MR effects in nanocomposite
particles are used in design of magneto-optical devices \cite{Raap}; the superposition principle for the
imaginary part of complex magnetic susceptibility of composite
magnetic fluids is crucial for the design of
absorbers and microwave attenuators which are based on ferromagnetic resonance
absorption of the electromagnetic field \cite {Fannin}.

Recently, Erdem investigated the magnetic relaxation in a spin-$1$ Ising model near the second-order phase
transition point.  In this study, time derivatives of the dipolar and quadrupolar
order parameters are treated as fluxes conjugate to their appropriate generalized forces in the sense of
irreversible thermodynamics \cite{Erdem}. In addition, Erdem has obtained the  frequency dependence of the complex susceptibility for the same system \cite{Erdemfreq}. Though the MR has been subject of many theoretical and experimental investigations mentioned above, there has been no
study to investigate the behavior of ac susceptibility in the metamagnetic Ising system.

Metamagnets, systems in which antiferromagnetic and ferromagnetic
 interactions exist simultaneously,
are of great interest because it is possible to
induce novel kinds of critical behavior by forcing
competition between these couplings, in particular
by applying a magnetic field \cite{Stryjewski}.
The {metamagnetic} model that has  in-plane ferromagnetic coupling and antiferromagnetic
coupling between adjacent planes, and the next nearest neighbor $(nnn)$ spin-$1/2$ Ising model with antiferromagnetic nearest-neighbour $(nn)$ and ferromagnetic nnn interactions  are the theoretical
Hamiltonian models showing a similar kind of behavior.
These models have been investigated by Monte-Carlo simulations
\cite{Landau}, as well as High-temperature series expansion calculations \cite{Harbus}.

$FeBr_{2}$ and $FeCl_{2}$ are typical Ising type metamagnets
\cite{Stryjewski}. In these structures, in the antiferromagnetic
phase when the iron ions in the triangular layers order
ferromagnetically, a layer with a negative sign follows a layer
with a positive sign. Due to this fact, the external field acts
differently on the oppositely oriented layers which  leads to
different ordered states and associates a  sequence of phase
transitions as a function of these two interaction strengths.
A Monte-Carlo simulation has been performed on a
realistic model of $FeCl_{2}$ under an external magnetic field
\cite{Hernandez}, in addition this typical metamagnet has also been treated by
a high-density expansion method on a two-sublattice
collinear Heisenberg Ising  model with
three- and four-ion anisotropy \cite{Onyszkiewicz1,Onyszkiewicz2}. Dynamic properties of  these systems have been
investigated by using three-spin flip
dynamics \cite{das}, Kawasaki dynamics \cite{sahni},  Glauber dynamics \cite{yan}, and  dynamic Monte Carlo renormalization group method \cite%
{oguz}. In our recent works we have calculated the kinetic phase diagrams  of the system under an oscillating
field \cite{gulphysicA,gulPLAglauber}. Moreover, we have presented an investigation of the relaxation dynamics of iron group dihalides and studied the field and temperature
dependence of the relaxation times via the phenomenological
kinetic coefficients \cite{GulPrevE}. Recently Gulpinar and Karaaslan  performed
an investigation of the relaxation times with taking in account the interference between the relaxation processes
of antiferromagnetic and ferromagnetic order parameters \cite{GulYenal}.

The purpose of this paper is to study the dynamical
magnetic response properties of the spin-$1/2$ Ising model in the presence of oscillating external
magnetic field.
On the other hand, to the best our knowledge there is no study in the literature which represents the mean field
calculations for the static
staggered and direct susceptibilities of a metamagnetic
Ising system.  Since then we have given the derivation of the static magnetic response functions and their temperature variations near the critical point in Sec.2.  Then, in Sec.3  we derived the temperature and frequency dependent
dynamic staggered and direct susceptibilities by a similar method previously
used to study relaxation dynamics and sound propagation \cite{GulAtenuasyon} and
investigate the behaviors of magnetic dispersion and absorption
factors near the second order phase transition temperature. The method used
in this paper provides information about  the
dynamical critical properties based on the phenomenological
kinetic coefficients  of the phenomenological rate
equations, which governs the magnetization relaxation, is due to an "ad
hoc" spin-lattice coupling.
%**************************************************************************************************************************
%**************************************************************************************************************************
\section{ Derivation of Static Staggered Magnetic and Magnetic Susceptibilities of Spin-1/2  Metamagnetic Ising  Model}
\hspace{0.5cm}
In order to obtain static staggered magnetic susceptibility one should introduce a staggered external field $B_{s}$ to the system \cite{Cohen}
whereas magnetic susceptibility is the response of the total magnetization to a physical external field $B$.
Since then one should add two different magnetic fields to the Hamiltonian of the spin-1/2 Metamagnetic Ising  Model:
\begin{equation}\label{1}
   \displaystyle \hat{H}=\displaystyle -\mathcal{J}\sum_{nn}S_{i}S_{j}+\mathcal{J}^{'}\sum_{nnn}S_{i}S_{j}-\mu B\sum_{i=1}^{N}S_{i}-\mu B_{s}
   \left( \sum_{1}S_{i}-\sum_{2}S_{j}\right).
\end{equation} $\mathcal{J}, \mathcal{J}^{'}, \mu, B$ and  $B_{s}$ are
ferromagnetic and antiferromagnetic exchange interaction constants, spin magnetic moment,
external magnetic field and external staggered field respectively.
By making use of mean field approximation  free energy of the system can be obtained  as following:
\begin{equation}
\label{2}
\begin{array}{lcl}
G(m_{t}, m_{s}, T, B, B_{s})&=&
\frac{1}{2}JN(m_{t}^{2}-m_{s}^{2})-\frac{1}{2}J^{'}N(m_{s}^{2}+m_{t}^{2})
\\
&  &-\frac{1}{4}k_{B}TN(4\ln(2)-(1+m_{s}+m_{t})\ln(1+m_{s}+m{t})
\\
& &-(1-m_{s}-m_{t})\ln(1-m_{s}-m_{t})-(1+m_{t}-m_{s})
\\
& &\ln(1+m_{t}-m_{s})-(1-m_{t}+m_{s})\ln(1-m_{t}+m_{s}))
\\
&& -Ng\mu _{B} B m_{t}-Ng\mu _{B} m_{s}B_{s},
\end{array}
\end{equation}
where $k_{B},g,\mu _{B},N$ are the Boltzmann's constant, the spin factor, the Bohr magneton and
the total number of metamagnetic Ising spins respectively.
 For a simple cubic lattice in which intralayer interactions are ferromagnetic and interlayer interactions  are antiferromagnetic $z=2$ and $z^{'}=4$.
 The equilibrium conditions,  $\left(\displaystyle \frac{\partial G}{\partial m_{t}}=0,\frac{\partial G}{\partial m_{s}}=0\right)$  result in the following transcendental equations:
\begin{equation}
\label{3}
\begin{array}{lcl}
\displaystyle m_{t} &=&\displaystyle \frac{\sinh(2\frac{B-m_{t}f}{k_{B}T})}{\cosh(2\frac{B-m_{t}f}{k_{B}T})+\cosh(2\frac{B_{s}+m_{s}t}{k_{B}T})},
\\
\\
\displaystyle m_{s} &=&\displaystyle \frac{\sinh(2\frac{B_{s}+m_{s}t}{k_{B}T})}{\cosh(2\frac{B-m_{t}f}{k_{B}T})+\cosh(2\frac{B_{s}+m_{s}t}{k_{B}T})}.
\end{array}
\end{equation}
Here, $f=J-J'$ and $t=J+J'$.
These equations may be solved without difficulty by an
iterative procedure, i.e. Newton Raphson method. The equilibrium
solutions should correspond to extremum of the
free energy so that one has to determine the solution that
minimizes $G$ \cite{Cohen}.
Since the solution of these equations are discussed in Ref.
\cite{GulPrevE} extensively, we shall only give a brief summary here as follows:
Topology of the metamagnetic Ising
model phase diagram depends on the value of the ratio of the
exchange interactions ($\eta=\frac{J}{J'}$):
(i)If $\eta >0.6$, the transitions between the anti-ferromagnetic and
paramagnetic phases are of first order at low temperatures and
strong fields while it is of second order at higher temperatures.
The two types of transitions are connected by a tricritical point.
(ii) For  $0< \eta < 0.6$, the tricritical point decomposes into
a critical end point (CEP) and a double cricital end point (DCP) with a line of first order transitions in between,
separating two anti-ferromagnetic phases \cite{Cohen,Selke}, see Fig.1(b)-(c) in Ref.\cite{GulPrevE}.
By definition staggered magnetic susceptibility is,
 \begin{equation}\label{4}
    \chi_{s_{static}}=\lim\limits_{B_{s}\rightarrow0}\frac{\partial m_{s}}{\partial B_{s}},\quad \quad (B=B_{0})
 \end{equation}
and total (direct) magnetic susceptibility can be expressed as,
  \begin{equation}\label{5}
\chi_{t_{static}}=\lim\limits_{B_{s}\rightarrow0}\frac{\partial m_{t}}{\partial B}\quad \quad (B=B_{0}).
\end{equation}
If one uses the  equations of state given in  Eq.(\ref{3}), the static  staggered magnetic and magnetic susceptibilities can be found as,
\begin{equation}\label{6}
\begin{array}{lcl}
  \chi_{s_{static}} & = & \displaystyle \lim\limits_{B_{s}\rightarrow0} \left(\frac{a_{21}c_{1}-a_{11}c_{2}}{a_{22}a_{11}-a_{21}a_{12}}\right),\\
  \\
  \chi_{t_{static}} & = & \displaystyle \lim\limits_{B_{s}\rightarrow0} \left(\frac{b_{22}d_{1}-b_{12}d_{2}}{b_{11}b_{22}-b_{12}b_{21}}\right).
\end{array}
\end{equation}\\Where, $a_{11}, a_{12}, a_{21}, a_{22}, c_{1},c_{2}$, and $b_{11}, b_{12}, b_{21}, b_{22}, d_{1},d_{2}$ are given in Appendix-A. \\
\indent {Fig.1(a)-(b) represents the critical behavior of static staggered magnetic susceptibility ($\chi_{s}$) and direct magnetic susceptibility. The arrows illustrate the phase transition temperature
$T_{C}(B)$ for $B=1.0$. The staggered magnetic susceptibility
($\chi_{s}$) increases rapidly with increasing temperature and
diverges as the temperature approaches to the second order phase
transition point on either side, as seen in Fig.1(a). On the other hand, the magnetic susceptibility
$(\chi_{t})$ also increases rapidly when the temperature is raised but
makes finite jump discontinuity  at the second order phase transition
temperature, which is illustrated in Fig.1(b).}
These findings are in accordance with the results given in  Refs.\cite{Zukovic1,Zukovic2}
obtained within  effective field approximation for an diluted metamagnetic Ising Model.
We should also note that Barry and Harrington have given an detailed analysis for the static total susceptibility of an antiferromagnetic Ising system in the Bethe-Takagi approximation which has the same result of the constant-coupling approximation of Kasteleyn and Van Kranendock applied to Ising antiferromagnetism \cite{Kasteleyn}.
In both studies an finite jump discontinuity is observed at the critical points which is in accordance with our results.
On the other hand, a  infinite-slope singularity is  found  by  exact series expansions  \cite{FisherandSykes}. One might also emphasize that  the mean field  approximation only barely fails to exhibit a maximum in the static  direct susceptibility above the $T_{N}$,  that implies a need for a method with takes into spin correlations \cite{barry}.

%**************************************************************************************************************************
%**************************************************************************************************************************
\section{{Derivation of Non-equilibrium Staggered Magnetic and Magnetic Susceptibilities of Spin-1/2  Metamagnetic Ising  Model}}
\vskip0.0cm \hskip1.1cm {In order to study the relaxation processes in the anhydrous iron group dihalides,
within  the spin-$\frac{1}{2}$ metamagnetic Ising model, one assumes  a small deviation in the value of the
applied external magnetic field.  This fact  removes the system slightly from its equilibrium state,
and one can investigate  how rapidly the metamagnetic system  relaxes back to its equilibrium
state. On the other hand, it is well known that the metamagnetic - paramagnetic phase transition lines
occur at places which are away from the $B=0$ axis \cite{Cohen}. Consequently, the magnetic Gibbs free energy
production ($\Delta {G} $) due to the deviations in the applied magnetic fields ($\delta B= B-B_{0}$, $\delta B_{s}= B_{s}-B_{s_{0}}$) can be expressed  as:}

\begin{equation}
\label{7}
G(m_{t}, m_{s}, T, B, B_{s})=G^{0}(m_{t_{0}}, m_{s_{0}}, T, B_{0}, B_{s_{0}})+\Delta G.
\end{equation}
{Where $G(m_{t},m_{s},T,B)$ is the free energy in the neighborhood of equilibrium, and  $G^{0}$ is the equilibrium Gibbs free energy and $\Delta G$ is the production of the Gibbs energy due to the variance of the external field, and is given following form:
}\begin{equation}
\label{8}
\begin{array}{lcl}
\Delta G&=&a(B-B_{0})+b(B_{s}-{B_{s_{0}}})+\frac{1}{2}c(m_{t}-m_{t_{0}})^{2}+\frac{1}{2}d(m_{s}-m_{s_{0}})^{2}
\\\\
 &    &+\frac{1}{2}e(B-B_{0})^{2}+\frac{1}{2}f(B_{s}-B_{s_{0}})^{2}+g(m_{t}-m_{t_{0}})(m_{s}-m_{s_{0}})
\\\\
 &    &+h(m_{t}-m_{t_{0}})(B-B_{0})+i(m_{t}-m_{t_{0}})(B_{s}-B_{s_{0}})
\\\\
 &    &+j(m_{s}-m_{s_{0}})(B-B_{0})+k(m_{s}-m_{s_{0}})(B_{s}-B_{s_{0}})
\end{array}
\end{equation}
{The coefficients $a-k$ are given in Appendix-B.}\\
\\
\indent {In the sense of Onsager's theory of irreversible thermodynamics, the time derivatives
of the antiferromagnetic and ferromagnetic order parameters are treated as generalized fluxes
conjugate to their appropriate generalized forces. One obtains the generalized forces
$(X_{m_{t}},X_{m_{s}})$ conjugate to the currents $(\overset{.}{m_{t}}, \overset{.}{m_{s}})$ respectively, by differentiating $\Delta G$ with
respect to  $(m_{t}-m_{t_{0}}),(m_{s}-m_{s0})$}:
\begin{equation}\label{9}
\begin{array}{lcl}
\displaystyle X_{m_{t}}=\displaystyle \frac{\partial \Delta G}{\partial (m_{t}-m_{t_{0}})} =c(m_{t}-m_{t_{0}})+g(m_{s}-m_{s_{0}})+h(B-B_{0})+i(B_{s}-B_{s_{0}} ),\\
\\
\\
\displaystyle  X_{m_{s}}=\displaystyle  \frac{\partial \Delta G}{\partial (m_{s}-m_{s_{0}})}= d(m_{s}-m_{s_{0}})+g(m_{t}-m_{t_{0}})+j(H-H_{0})+k(B_{s}-B_{s_{0}} ).
\end{array}
\end{equation}
{The linear relations between the currents and forces may
be written in terms of a matrix of phenomenological rate
coefficients and since both $m_{t}$ and $m_{s}$ are odd variables under time inversion
this matrix should be  symmetric \cite{groot}:
}
\begin{equation}\label{10}
\left[
\begin{array}{l}
\dot m_{t} \\
\dot m_{s}%
\end{array}%
\right] =\left[
\begin{array}{cc}
\gamma _{m_{t}} & \gamma  \\
\gamma  & \gamma _{_{m_{s}}}%
\end{array}%
\right] \left[
\begin{array}{c}
X_{m_{t}} \\
X_{m_{s}}%
\end{array}%
\right].
\end{equation}%
Consequently, this matrix equation can be written in component
form using equations (\ref{9}), namely a set of two coupled,
linear inhomogenous first order rate equations,

\begin{equation}\label{11}
\begin{array}{lcl}
\dot m_{t}&=& (\gamma_{m_{t}}c+\gamma g)(m_{t}-m_{t_{0}})+(\gamma_{m_{t}}g+\gamma d)(m_{s}-m_{s_{0}})
\\\\&&
+(\gamma_{m_{t}}h+\gamma j)(B-B_{0})+(\gamma_{m_{t}}i+\gamma k)(B_{s}-B_{s_{0}}),
\\\\
\dot m_{s}&=&(\gamma c+\gamma_{m_{s}}g)(m_{t}-m_{t_{0}})+(\gamma g+\gamma_{m_{s}}d)(m_{s}-m_{s_{0}})
\\\\
&&
+(\gamma h+\gamma_{m_{s}}j)(B-B_{0})+(\gamma i+\gamma_{m_{s}}k)(B_{s}-B_{s_{0}}).
\end{array}
\end{equation}
In order to find the relaxation times, one considers the
corresponding homogeneous equations resulting when there is
no external stimulation, namely $B=B_{0},B_{s}=B_{s_{0}} $. Eqs. (\ref{11}) then become
\begin{equation}\label{12}
\left[
\begin{array}{l}
\dot{m_{t}} \\
\dot{m_{s}}%111
\end{array}%
\right] =\left[
\begin{array}{cc}
\gamma_{m_{t}}c+\gamma g & \gamma_{m_{t}}g+\gamma d \\
\gamma c+\gamma_{m_{s}}g & \gamma g+\gamma_{m_{s}}d
\end{array}%
\right] \left[
\begin{array}{c}
m_{t}-m_{t_0} \\
m_{s}-m_{s_0}%
\end{array}%
\right],
\end{equation}
\indent {In iron group dihalides and all metamagnetic spin systems, there exists two
order parameters, total and staggered magnetization  ($m_{t}$ and $m_{s}$) which characterize the magnetic behavior of the system.
Further, one can see from  Eq.(\ref{2}) that, antiferromagnetic and ferromagnetic order are coupled
to each other. Due to this fact, there is an inteference between the ferromagnetic and antiferromagnetic relaxation processes. In the Onsager's theory  of irreversible thermodynamics, the effect of cross effects between these two relaxation processes in the iron group dihalides is embedded in the kinetic rate coefficient  $\gamma$. Further,  as it is discussed extensively by  Barry and Harrington in Ref.\cite{barry}, other operator quantities should be added to the Ising Hamiltonian which do not commute with  $\mathbf{\sigma}_{z}$ thereby introducing transitions within the spin system permitting, consequently, longitudinal relaxation. It is well known that, longitudinal relaxation is closely related to spin-lattice relaxation time so that these added operator quantities should contain some kind of spin-lattice coupling, e.g. the frequently designated $T_{1}$  appearing in the longitudinal Bloch equation representation of spin-lattice relaxation in solids \cite{barry}. \\
\\
\indent The kinetic equations of motions given by Eqs.(\ref{11}) can be solved by assuming a form for the solution
( $m_{t}-m_{t_{0}},m_{s}-m_{s_{0}}\simeq e^{-\frac{t}{\tau}}$ )
by making use of the secular equation given below
Thus, one obtains the following
secular equation:
                   \begin{equation}
                      {\left\vert
                      \begin{array}{cc}
                      \frac{1}{\tau} +\gamma _{m}a-\gamma b & \gamma c-\gamma_{m}b \\
                      \gamma a-\gamma _{m_{s}}b & \frac{1}{\tau}+\gamma _{m_{s}}c-\gamma b
                      \end{array}%
                      \right\vert }=0,
                      \label{13}
                      \end{equation}
the resulting relaxation times can be found  as:}
\begin{equation}\label{14}
 \begin{array}{lcl}
   \displaystyle\frac{1}{\tau_{1}}=-\gamma g-\frac{1}{2}\gamma _{m_{s}}d-\frac{1}{2}\gamma _{m_{t}}c+\frac{1}{2}\Delta^{1/2}
    \\
      \\
   \displaystyle\frac{1}{\tau_{2}}=-\gamma g-\frac{1}{2}\gamma _{m_{s}}d-\frac{1}{2}\gamma _{m_{t}}c-\frac{1}{2}\Delta^{1/2}
 \end{array}
\end{equation}
where $\Delta$ is
\begin{equation}\label{15}
\Delta=\left(4\gamma_{m_{s}}d\gamma g+4\gamma g\gamma_{m_{t}}c+\gamma_{m_{s}}^{2}d^{2}-2\gamma_{m_{t}}c\gamma_{m_{s}}d+\gamma_{m_{t}}^{2}c^{2}+4\gamma_{m_{t}}g^{2}\gamma_{m_{s}}+4\gamma^{2}dc\right).
\end{equation}
Temperature dependencies of the relaxation times for non zero external magnetic field
in the metamagnetic (antiferromagnetic) and paramagnetic phases for the
cases  ($\gamma _{m}\gamma _{m_{s}}-\gamma ^{2}\approx \gamma _{m}\gamma _{m_{s}}$) and  ($\gamma ^{2}\geq \gamma _{m}\gamma _{m_{s}}$)
were discussed in Refs. \cite{GulPrevE,GulYenal} .
%**************************************************************************************************************************
%**************************************************************************************************************************
\subsection{{ Derivation of Kinetic Equations Leading to AC Total Susceptibility}}
{If one stimulates the metamagnetic Ising system  by magnetic field B oscillating at an angular frequency $\omega$, all quantities will oscillate near the equilibrium state at this same angular frequency:
}
\begin{equation}\label{16}
  m_{t}-m_{t_{0}}=m_{t_{1}}e^{I\omega t}, \quad m_{s}-m_{s_{0}}=m_{s_{1}}e^{I\omega t}, \quad B-B_{0}=B_{1}e^{I\omega t}.
\end{equation}
{Substituting these equations into the kinetic equations given by  Eqs.(\ref{11}) for $B_{s}=B_{s_{0}}$ we find following form:  }
\begin{equation}\label{17}
  \begin{array}{lcl}
    \left(-I\omega+\gamma_{m_{t}}c+\gamma g  \right)m_{t_{1}}+\left( \gamma_{m_{t}}g+\gamma d\right)m_{s_{1}}&=&-\gamma_{m_{t}}hB_{1},
    \\
    \\
    \left( \gamma c+\gamma_{m_{s}}g\right)m_{t_{1}}+\left( -I\omega+\gamma g+\gamma_{m_{s}}d\right)m_{s_{1}}&=&-\gamma hB_{1}.
  \end{array}
\end{equation}
{After some algebra, we can find the matrix form of  $\displaystyle\frac{m_{t_{1}}}{B_{1}}$:}
\begin{equation}\label{18}
\frac{m_{t_{1}}}{B_{1}}=\frac{\left|
                          \begin{array}{cc}
                            -\gamma_{m_{t}}h & \gamma_{m_{t}}g+\gamma d \\
                            -\gamma h & -I\omega+\gamma g+\gamma_{m_{s}}d \\
                          \end{array}
                        \right|}{\left|
                                  \begin{array}{cc}
                                    -I\omega+\gamma_{m_{t}}c+\gamma g  & \gamma_{m_{t}}g+\gamma d \\
                                     \gamma c+\gamma_{m_{s}}g & -I\omega+\gamma g+\gamma_{m_{s}}d \\
                                  \end{array}
                                \right|}.
\end{equation}
The determinant in the denominator of Eq. (20) is the same as the
secular determinant used to calculate the reciprocal relaxation
times given by Eq.(\ref{13}) except for the replacement of $\frac{1}{\tau}$ by $i \omega$. Hence
Eq. (20) may be written as
\begin{equation}\label{19}
\frac{m_{t_{1}}}{B_{1}}=-h\frac{\left(\gamma_{m_{t}}(-I\omega)+d(\gamma_{m_{t}}\gamma_{m_{s}}-\Gamma^{2})\right)}{(-I\omega-\tau_{1}^{-1})(-I\omega-\tau_{2}^{-1})}.
\end{equation}
Eq.(\ref{19}) can be used in order to calculate the complex total
magnetic susceptibility $\chi_{t}(\omega)$. This may be seen as follows: The
induced magnetization (total induced magnetic
moment per unit volume) of the spin-$\frac{1}{2}$ Ising model with $nn$ and $nnn$ interactions   is given by
\begin{equation}\label{20}
  m_{t}-m_{t_{\infty}}=Re(m_{t_{1}}e^{I\omega t}),
\end{equation}
where $m_{t_{\infty}}$ is the magnetization induced by a magnetic field
oscillating at $\omega$. Also, by definition, the expression for $\chi_{t}(\omega)$
may be written
\begin{equation}\label{21}
   m_{t}-m_{t_{\infty}}=Re\left(\chi_{t}(\omega)B_{1}e^{I\omega t}\right),
\end{equation}
where
\begin{equation}\label{22}
\chi_{t}(\omega)= \chi_{t}^{'}(\omega)- i \chi_{{t}}^{''}(\omega)
\end{equation}
here  $\chi_{m_{s}}^{'}(\omega)$, $\chi_{m_{s}}^{''}(\omega)$ are the magnetic
dispersion and absorption factor, respectively. Comparing
Eqs. (\ref{21}) and (\ref{22}), one may write
\begin{equation}\label{23}
\chi_{t}(\omega)=\frac{m_{t_{1}}}{B_{1}},
\end{equation}
and
\begin{equation}\label{24}
\chi'_{t}(\omega)=h\frac{(\tau_{1}\tau_{2})}{(\tau_{2}-\tau_{1})}\left(\frac{\gamma_{m_{t}}+\tau_{1}d(\gamma_{m_{s}}\gamma_{m_{t}}-\Gamma^{2})}
{1+\omega^{2}\tau_{1}^{2}}-\frac{\gamma_{m_{t}}+\tau_{2}d(\gamma_{m_{s}}\gamma_{m_{t}}-\Gamma^{2})}{1+\omega^{2}\tau_{2}^{2}}\right),
\end{equation}

\begin{equation}\label{25}
\chi''_{t}(\omega)=h\frac{(\tau_{1}\tau_{2})}{(\tau_{2}-\tau_{1})}
\left(\frac{\omega\tau_{1}\left(\gamma_{m_{t}}+\tau_{1}d(\gamma_{m_{s}}\gamma_{m_{t}}-\Gamma^{2}\right))}
{1+\omega^{2}\tau_{1}^{2}}
-\frac{\omega\tau_{2}(\gamma_{m_{t}}+\tau_{2}d(\gamma_{m_{s}}\gamma_{m_{t}}-\Gamma^{2}))}{1+\omega^{2}\tau_{2}^{2}}\right).
\end{equation}

%**************************************************************************************************************************
%**************************************************************************************************************************
\subsection{{ Derivation of Kinetic Equations Leading to Dynamical Staggered Magnetic Susceptibility}}
Now we assume that  the system  is stimulated   by a staggered magnetic field $B_{s}$ oscillating at an angular frequency $\omega$, all quantities will oscillate near the equilibrium state at this angular frequency:
\begin{equation}\label{26}
    m_{t}-m_{t_{0}}=m_{t_{1}}e^{I\omega t}, \quad m_{s}-m_{s_{0}}=m_{s_{1}}e^{I\omega t}, \quad B_{s}-B_{s_{0}}=B_{s_{1}}e^{I\omega t},
\end{equation}
{embedding Eqs.(\ref{26}) into the kinetic equations given by  Eqs.(\ref{11}) for $B=B_{0}$ one obtains  }
\begin{equation}\label{27}
\begin{array}{lcl}
    \left(-I\omega+\gamma_{m_{t}}c+\gamma g  \right)m_{t_{1}}+\left( \gamma_{m_{t}}g+\gamma d\right)m_{s_{1}}&=&-\gamma k B_{s_{1}},
    \\
    \\
    \left( \gamma c+\gamma_{m_{s}}g\right)m_{t_{1}}+\left( -I\omega+\gamma g+\gamma_{m_{s}}d\right)m_{s_{1}}&=&-\gamma_{m_{s}} k B_{s_{1}}.
  \end{array}
\end{equation}
solving the Eq(\ref{27} ) yields the following  matrix form of  $\displaystyle\frac{m_{s_{1}}}{B_{s_{1}}}$
\begin{equation}\label{28}
  \frac{m_{s_{1}}}{B_{s_{1}}}= \frac{\left|
                                      \begin{array}{cc}
                                        -I\omega+\gamma_{m_{t}}c+\gamma g  & -\gamma k  \\
                                        \gamma c+\gamma_{m_{s}}g & -\gamma_{m_{s}} k  \\
                                      \end{array}
                                    \right|}{\left|
                                 \begin{array}{cc}
                                   -I\omega+\gamma_{m_{t}}c+\gamma g   & \gamma_{m_{t}}g+\gamma d  \\
                                   \gamma c+\gamma_{m_{s}}g &-I\omega+\gamma g+\gamma_{m_{s}}d\\
                                 \end{array}
                               \right|}.
\end{equation}
Following the same analysis given in Sec.3.1 one obtains
\begin{equation}\label{29}
 \frac{m_{s_{1}}}{B_{s_{1}}}=-k\frac{\left(\gamma_{m_{s}}(-I\omega)+c(\gamma_{m_{t}}\gamma_{m_{s}}-\Gamma^{2})\right)}{(-I\omega-\tau_{1}^{-1})(-I\omega-\tau_{2}^{-1})},
\end{equation}
The induced staggered magnetization due to the effect of the oscillating staggered magnetic field is
\begin{equation}\label{30}
m_{s}-m_{s_{\infty}}=Re(m_{s_{1}}e^{I\omega t}).
\end{equation}
where $m_{s_{\infty}}$ is the staggered magnetization induced by a stagerred field oscillating at infinite frequency.
Also, by definition the expression for   the complex   stagerred susceptibility reads
\begin{equation}\label{31}
m_{s}-m_{s_{\infty}}=Re(\chi_{s}(\omega)B_{s_{1}}e^{I\omega t}).
\end{equation}
By making use of Eqs.(\ref{28}) and (\ref{31}) one obtains
\begin{equation}\label{32}
 \chi_{s}(\omega)=\frac{m_{s_{1}}}{B_{s_{1}}},
\end{equation}
finally,  by using the relation $\chi_{{s}}(\omega)= \chi_{{s}}^{'}(\omega)- i \chi_{{s}}^{''}(\omega)$
we find
\begin{equation}\label{33}
\chi'_{s}(\omega)=k\frac{(\tau_{1}\tau_{2})}{(\tau_{2}-\tau_{1})}\left(\frac{\gamma_{m_{s}}+\tau_{1}c(\gamma_{m_{s}}\gamma_{m_{t}}-\Gamma^{2})}{1+\omega^{2}\tau_{1}^{2}}-\frac{\gamma_{m_{s}}+\tau_{2}c(\gamma_{m_{s}}\gamma_{m_{t}}-\Gamma^{2})}{1+\omega^{2}\tau_{2}^{2}}\right),
\end{equation}
\begin{equation}\label{34}
\chi_{s}^{''}(\omega)=k\frac{(\tau_{1}\tau_{2})}{(\tau_{2}-\tau_{1})}\left(\frac{\omega\tau_{1}(\gamma_{m}+\tau_{1}c(\gamma_{m_{s}}\gamma_{m_{t}})-\Gamma^{2})}{1+\omega^{2}\tau_{1}^{2}}-\frac{\omega\tau_{2}(\gamma_{m}+\tau_{2}c(\gamma_{m_{s}}\gamma_{m_{t}}-\Gamma^{2})}{1+\omega^{2}\tau_{2}^{2}}\right).
\end{equation}
here $\chi_{{s}}^{'}(\omega)$, $\chi_{{s}}^{''}(\omega)$ are the magnetic dispersion and absorption factors respectively.

%**************************************************************************************************************************
%**************************************************************************************************************************
\section{Results}

{Fig.2(a)-(b) shows the critical behavior of  staggered magnetic dispersion, $\chi_{s}^{'}(\omega)$ and direct magnetic dispersion $\chi_{t}^{'}(\omega)$ with varying temperature in the low frequency region $\omega \tau_{2}\ll1$ . }
One can observe from Fig.2(a) that real part of the staggered ac susceptibility ({$\chi_{s}^{'}(\omega)$)
rises rapidly with increasing temperature and  tends to infinity near the second order phase transition.
In accordance with the expectations the staggered dispersion factor converges to static staggered susceptibility for  $\omega \tau_{2}\ll1$ (compare Fig.1(a) and Fig.2(a)). Fig2.(b) represents the temperature variation of the direct magnetic dispersion factor in the neighborhood of $T_{N}$ for $B=1.0$ and $\eta=2.0$. The real part of direct complex susceptibility increases slowly with changing temperature and makes a cusp behavior near the second order phase transition.
 In  addition,  $\chi_{t}^{'}(\omega)$ converges to static direct susceptibility in the low frequency region (compare Fig.2(b) and Fig.1(b)).
 Moreover the behavior of $\chi_{s}^{'}(\omega)$ and $\chi_{t}^{'}(\omega)$ are independent of the frequency}. This behavior is in accordance with results obtained for spin-1 Ising model \cite{Erdem}.

{\indent Fig. 3(a)-(b) illustrates temperature variations of staggered magnetic dispersion  and direct magnetic
dispersion  for the high frequency region at several values of $\omega$ in the neighborhood second phase transition point. $\chi_{s}^{'}(\omega)$ has  a local maximum before the second order phase transition point and shows a local
minimum near the critical point in the high frequency region. In these figures the number accompanying
each curve denotes the value of o and the vertical arrows refer to the critical temperature for $B=1.0$ and $\eta=2.0$.
The direct dispersion factor exhibits a maxima for $T\rightarrow T_{N}$ whose amplitude increases with rising frequency.
This result is in parallel  with findings for the total magnetic dispersion factor for the antiferromagnetic Ising model \cite{barry}.

}{ In Fig.4(a)-(b) temperature variations of staggered magnetic absorption  and direct magnetic absorption  are given for the low frequency region at several values of $\omega$. $\chi_{s}^{''}$ exhibits a divergence, while  $\chi_{t}^{''}$ makes a local maximum near the critical point. Fig.5(a)-(b) shows  temperature variations of staggered magnetic absorption  and direct magnetic absorption  for the high frequency region at several values of $\omega$ in the neighborhood second phase transition point.  $\chi_{s}^{''}$ exhibits a local maximum at the phase transition point  and amplitude  of the maximum changes with frequency (See Fig. 5(a)). On the other hand, $\chi_{t}^{''}$ makes local minimum at the phase transition point, but it exhibits  a local maximum before the critical temperature (See Fig. 5(b)).}
The divergencies of $\chi_{s}^{'}(\omega)$ and $\chi_{s}^{''}(\omega)$ for $\omega \tau_{2}\ll1$  on both sides of the critical
point are characterized by the critical exponents. Thus, we may
assume that, for temperatures smaller than the transition
temperatures, $\chi_{s}^{'}(\omega)$ and $\chi_{s}^{''}(\omega)$ follow laws of the form $\chi_{s}^{'}(\omega)\sim |T-T_{N}|^{\gamma'_{ac}}$
and $\chi_{s}^{'}(\omega)\sim |T-T_{N}|^{\gamma''_{ac}}$, respectively.
Here $\gamma'_{ac}$ and $\gamma''_{ac}$ are the dynamic critical exponents of the staggered magnetic dispersion and absorption factors.
In order to calculate the mean field values of $\gamma'_{ac}$ and $\gamma''_{ac}$, we have sketched the
the plots of $\chi_{s}^{'}(\omega)$ versus $log(1-\frac{T}{T_{N}})$ and
$\chi_{s}^{'}(\omega)$ versus $log(1-\frac{T}{T_{N}})$ which are represented in Fig.6 (a).
We have found only one linear part on these $log-log$ plots for $\chi_{s}^{'}(\omega)$ and $\chi_{s}^{''}(\omega)$, i.e., one values of $\gamma'_{ac}$ and $\gamma''_{ac}$, respectively.
For each $\omega$ value the slope of the line is equal to $-1$ for
$\gamma'_{ac}$; and $-2$ for $\gamma''_{ac}$.
These values are in accordance with the diverging behavior of the magnetic dispersion and absorption factor near the second order phase transition
temperature for $B=B_{0}$ and $B_{s}=B_{s_{0}}=0$.
It is important to note that these findings are agreement with the critical exponents found for the magnetic
spin-1 Ising model \cite{Erdem}.
Further these results are in well agreement with the MR studies of an AB type Ising model for  for Ising antiferromagnetism \cite{barry} and Ising ferromagnet \cite{barry_F}.In addition, Fig.6(b) and (c) show logarithmic plot of staggered magnetic dispersion and staggered magnetic
absorption versus reduced temperature $(1-T/T_{C})$ for $T < T_{C}$ at several values in the
low frequency region  and high frequency region. Finally, Table-1 represents mean field critical exponents for  staggered magnetic dispersion, staggered magnetic absorbtion, direct magnetic dispersion  and  direct magnetic absorbtion in the low and high frequency regions.
\section{Concluding Remarks}

Within the mean field approximation, we have analyzed steady
state solutions of the spin-$\frac{1}{2}$ metamagnetic Ising model under
a time-dependent oscillating external physical and staggered magnetic fields.
In this paper, the formulation is based on a method  which combines the  equilibrium statistical theory of critical phenomena  with the theory of irreversible
thermodynamics. It is assumed that the amplitude of both the physical and staggered fields are so small that
we made use of linear response theory in studying the magnetic relaxation processes
in a metamagnetic system. In this system there exists two coupled
relaxation processes which correspond to the relaxation of
ferromagnetic and antiferromagnetic order parameters. We have shown that these processes
are characterized by two distinct relaxation times ($\tau_{1}$ and $\tau_{2}$ ).
$\tau_{2}$ is the dominant relaxation time which characterizes the critical
slowing down of the staggered magnetization and therefore the temperature variation of the second relaxation time
determines the separation of the so-called low- and high-frequency
regions. We should note that  similar behavior has been observed in
the investigation of the antiferromagnetic Ising model with the same method \cite{barry}.
Since $\frac{1}{\tau_{2}}\rightarrow 0$ as $T \rightarrow T_{N}$ and keeping the frequency $\omega$
fixed, we observed the low-frequency behaviors followed by
the high-frequency behaviors for the dynamic susceptibilities.
One can see from Figs.2(a) and 3(a) that
the slope of the staggered magnetic dispersion curve
chances in sign as $T \rightarrow T_{N}$ (positive slope for
the low-frequency region, negative slope for the high-frequency
region). Similar behavior exists also in the high and low frequency regions for total magnetic dispersion factor.
Finally, it should be emphasized that we have assumed in this study
the rate coefficients have negligible temperature dependence. The
validity of this assumption should be testified
either by experiments or a more powerful  theory such as path probability method.
Kikuchi has represented an investigation of
Order-disorder configurational relaxation on a bcc  AB-type lattice and showed that
The diagonal Onsager coefficients tend to finite values whereas the off-diagonal
coefficient which characterizes the interference between coupled irreversible processes \cite{Onsager}
tends to vanish as temperature approaches critical temperature \cite{Kikuchi}.

%**************************************************************************************************************************
%**************************************************************************************************************************
\section{Acknowledgements}
This work was supported by the Scientific and Technological Research
Council of Turkey (TUBITAK), Grant No. 109T721. In addition authors thank
A.N. Berker for valuable discussions, Sabanci University and Massachusetts
Institute of Technology%**************************************************************************************************************************
%**************************************************************************************************************************
\section{Appendix-A}
\noindent The list of coefficients in Eqs.(\ref{6}):
\begin{equation}\label{katsayilar}
\begin{array}{lcl}
a_{11}&=&\displaystyle 1+\frac{2\cosh(2\frac{B-m_{t}f}{k_{B}T})f}{k_{B}T\left(\cosh(2\frac{B-m_{t}f}{k_{B}T})+\cosh(2\frac{B_{s}+m_{s}t}{k_{B}T})\right)}-\frac{2\sinh(2\frac{B-m_{t}f}{k_{B}T})^2f}{k_{B}T\left(\cosh(2\frac{B-m_{t}f}{k_{B}T})+\cosh(2\frac{B_{s}+m_{s}t}{k_{B}T})\right)^2},
\\
\\
a_{12}&=&\displaystyle \frac{2\sinh(2\frac{B-m_{t}f)}{k_{B}T})\sinh(2\frac{B_{s}+m_{s}t)}{k_{B}T})t}{k_{B}T\left(\cosh(2\frac{B-m_{t}f}{k_{B}T})+\cosh(2\frac{B_{s}+m_{s}t}{k_{B}T})\right)^2},
\\
\\
a_{21}&=&\displaystyle-\frac{2\sinh(2\frac{B_{s}+m_{s}t}{k_{B}T})\sinh(2\frac{B-m_{t}f}{k_{B}T})f}{k_{B}T\left(\cosh(2\frac{B-m_{t}f}{k_{B}T})+\cosh(2\frac{B_{s}+m_{s}t}{k_{B}T})\right)^{2}},
\\
\\
\end{array}
\end{equation}
$$
\begin{array}{lcl}
a_{22}&=&1-\displaystyle\frac{2\cosh(2\frac{B_{s}+m_{s}t}{k_{B}T})t}{k_{B}T\left(\cosh(2\frac{B-m_{t}f}{k_{B}T})+\cosh(2\frac{B_{s}+m_{s}t}{k_{B}T})\right)}+\frac{2\sinh(2\frac{B_{s}+m_{s}t}{k_{B}T})^{2}t}{k_{B}T\left(\cosh(2\frac{B-m_{t}f}{k_{B}T})+\cosh(2\frac{B_{s}+m_{s}t}{k_{B}T})\right)^{2}},
\\
\\
c_{1}&=&-\displaystyle\frac{2\sinh(2\frac{B-m_{t}f}{k_{B}T})\sinh(2\frac{B_{s}+m_{s}t}{k_{B}T})}{k_{B}T\left(\cosh(2\frac{B-m_{t}f}{k_{B}T})+\cosh(2\frac{B_{s}+m_{s}t}{k_{B}T})\right)^{2}},
\\
\\
c_{2}&=&\displaystyle \frac{2\cosh(2\frac{B_{s}+m_{s}t}{k_{B}T})}{k_{B}T\left(\cosh(2\frac{B-m_{t}f}{k_{B}T})+\cosh(2\frac{B_{s}+m_{s}t}{k_{B}T})\right)}-\frac{2\sinh(2\frac{B_{s}+m_{s}t}{k_{B}T})^{2}}{k_{B}T\left(\cosh(2\frac{B-m_{t}f}{k_{B}T})+\cosh(2\frac{B_{s}+m_{s}t}{k_{B}T})\right)^{2}}.
\end{array}
$$
\begin{equation}\label{katsayilar2}
\begin{array}{lcl}
b_{11}&=& 1+\displaystyle\frac{2\cosh(2\frac{B-m_{t}f)}{k_{B}T})f}{k_{B}T\left(\cosh(2\frac{B-m_{t}f}{k_{B}T})+\cosh(2\frac{B_{s}s+m_{s}t}{k_{B}T})\right)}-\frac{2\sinh(2\frac{B-m_{t}f}{k_{B}T})^{2}f}{\left(\cosh(2\frac{B-m_{t}f}{k_{B}T})+\cosh(2\frac{B_{s}+m_{s}t}{k_{B}T})\right)^{2}k_{B}T},
\\
\\
b_{12}&=&\displaystyle \frac{2\sinh(2\frac{B-m_{t}f}{k_{B}T})\sinh(2\frac{B_{s}+m_{s}t}{k_{B}T})t}{\left(\cosh(2\frac{B-m_{t}f}{k_{B}T})+\cosh(2\frac{B_{s}+m_{s}t}{k_{B}T})\right)^{2}k_{B}T},
\\
\\
b_{21}&=&-\displaystyle \frac{2\sinh(2\frac{B_{s}+m_{s}t}{k_{B}T})(\sinh(2\frac{B-m_{t})f}{k_{B}T})f}{\left(\cosh(2\frac{B-m_{t}f}{k_{B}T})+\cosh(2\frac{B_{s}s+m_{s}t}{T})\right)^{2}k_{B}T},
\\
\\
\end{array}
\end{equation}
\begin{equation}
\begin{array}{lcl}
b_{22}&=&1-\displaystyle \frac{2\cosh(2\frac{B_{s}+m_{s}t}{k_{B}T})t}{k_{B}T\left(\cosh(2\frac{B-m_{t}f}{k_{B}T})+\cosh(2\frac{B_{s}+m_{s}t}{k_{B}T})\right)}+\frac{2\sinh(2\frac{B_{s}+m_{s}t)}{k_{B}T})^{2}t}{\left(\cosh(2\frac{B-m_{t}f}{k_{B}T})+\cosh(2\frac{B_{s}+m_{s}t}{k_{B}T})\right)^{2}k_{B}T},
\\
\\
d_{1}&=&\displaystyle \frac{2\cosh(2\frac{B-m_{t}f}{k_{B}T})}{k_{B}T\left(\cosh(2\frac{B-m_{t}f}{k_{B}T})+\cosh(2\frac{B_{s}+m_{s}t}{k_{B}T})\right)}-\frac{2\sinh(2\frac{B-m_{t}f}{k_{B}T})^{2}}{k_{B}T\left(\cosh(2\frac{B-m_{t}f}{k_{B}T})+\cosh(2\frac{B_{s}+m_{s}t}{k_{B}T})\right)^{2}},
\\
\\
d_{2}&=&-\displaystyle \frac{2\sinh(2\frac{B_{s}+m_{s}t}{k_{B}T})\sinh(2\frac{B-m_{t}f}{k_{B}T})}{\left(\cosh(2\frac{B-m_{t}f}{k_{B}T})+\cosh(2\frac{B_{s}+m_{s}t}{k_{B}T})\right)^{2}k_{B}T}.
\end{array}
\end{equation}
%**************************************************************************************************************************
%**************************************************************************************************************************
\section{Appendix-B}
The list of coefficients in Eq.(8):
$$
\begin{array}{lcl}
a  & =& \displaystyle \left(\frac{\partial G}{\partial B} \right)_{eq}=-N\mu m_{t}
\\
\\
b  & =& \displaystyle \left(\frac{\partial G}{\partial B_{s}} \right)_{eq}=-N\mu m_{s}
\\
\\
c  & =& \displaystyle \left(\frac{\partial^{2} G}{\partial m_{t}^{2} } \right)_{eq},
\\
\\
&=&\displaystyle N\left( J-J^{'}-\frac{1}{4}k_{B}T\left(-\frac{1}{1+m_{s}+m_{t}}-\frac{1}{1-m_{s}-m_{t}}-\frac{1}{1+m_{t}-m_{s}}-\frac{1}{1-m_{t}+m_{s}}\right)\right)
\\
\\
d  & =& \displaystyle \left(\frac{\partial^{2} G}{\partial m_{s}^{2} } \right)_{eq},
\\
\\
& = & \displaystyle -N\left(J+J^{'}+\frac{1}{4}k_{B}T\left(-\frac{1}{1+m_{s}+m_{t}}-\frac{1}{1-m_{s}-m_{t}}-\frac{1}{1+m_{t}-m_{s}}-\frac{1}{1-m_{t}+m_{s}}\right)\right)
\\
\\
e  & =& \displaystyle \left(\frac{\partial^{2} G}{\partial B^{2} } \right)_{eq}=0,
\\
\\
f  & =& \displaystyle \left(\frac{\partial^{2} G}{\partial B_{s}^{2} } \right)_{eq}=0,
\\
\\
g  & =& \displaystyle \left(\frac{\partial^{2} G}{\partial m_{t}\partial m_{s} } \right)_{eq},
\\
\\
&=& \displaystyle -\frac{1}{4}k_{B}TN\left(-\frac{1}{1+m_{s}+m_{t}}-\frac{1}{1-m_{s}-m_{t}}+\frac{1}{1+m_{t}-m_{s}}+\frac{1}{1-m_{t}+m_{s}}\right),
\\
\\
h  & =& \displaystyle \left(\frac{\partial^{2} G}{\partial B \partial m_{t} } \right)_{eq}=-N\mu,
\\
\\
i  & =& \displaystyle \left(\frac{\partial^{2} G}{\partial m_{t}\partial B_{s} } \right)_{eq}=0,
\\
\\
j  & =& \displaystyle \left(\frac{\partial^{2} G}{\partial B \partial m_{s} } \right)_{eq}=0,
\\
\\
k  & =& \displaystyle \left(\frac{\partial^{2} G}{\partial B_{s}\partial m_{s} } \right)_{eq}=-N\mu.
\end{array}
$$
%**************************************************************************************************************************
%**************************************************************************************************************************

%**************************************************************************************************************************
%**************************************************************************************************************************
\newpage
\begin{figure}[tbp]
\begin{center}
 \includegraphics[width=6cm,height=7cm,angle=0,bb=0 0 1000 1000]{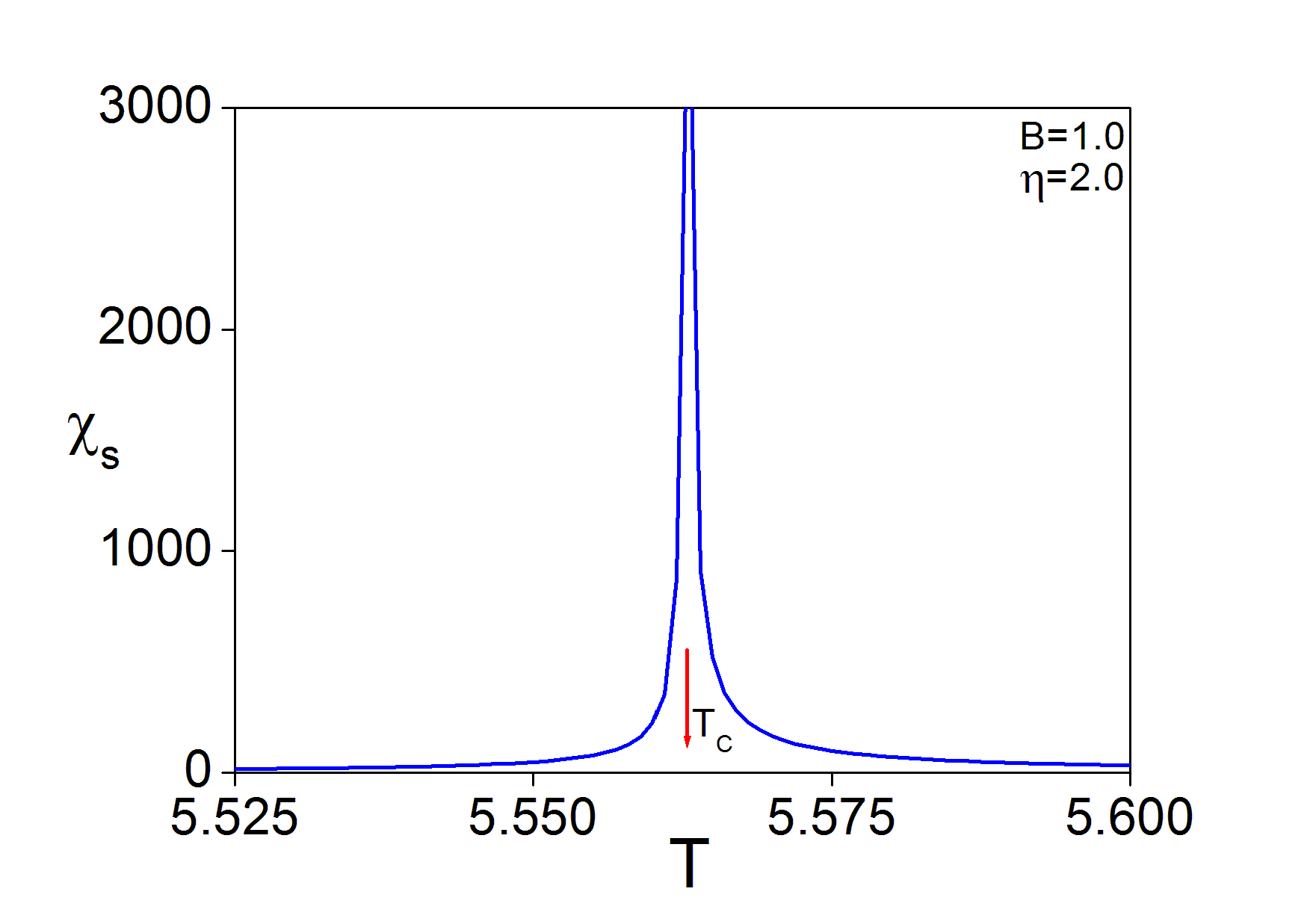}
  \includegraphics[width=6cm,height=7cm,angle=0,bb=0 0 1000 1000]{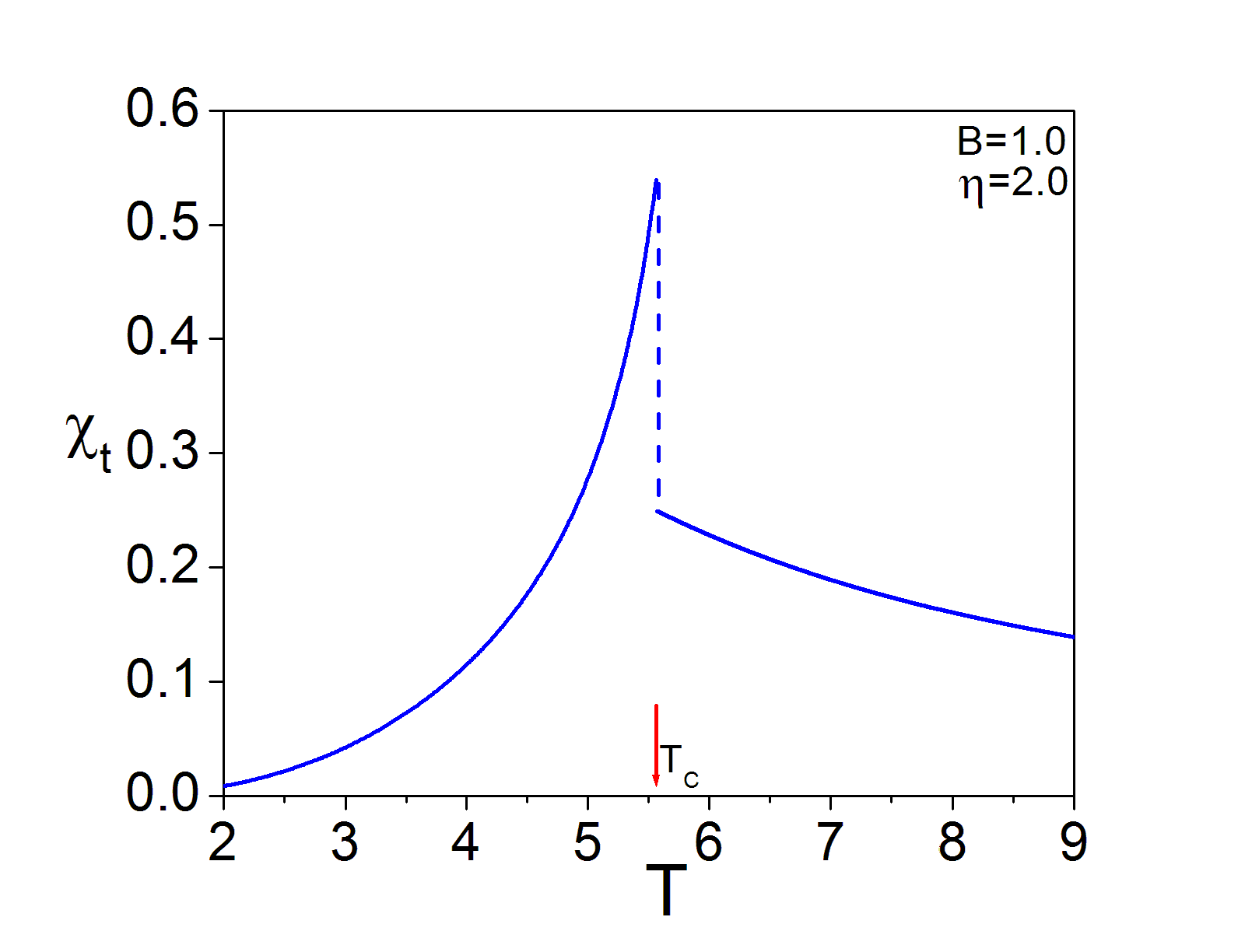}
 \caption{Constant field (B) cross-sections of the  staggered magnetic susceptibility (a) and direct susceptibility (b) in the neighborhood of  critical point while $B=1.0$ and  $\eta=\frac{zJ}{z'J'}$.}
\end{center}
\end{figure}

\begin{figure}[tbp]
\begin{center}
 \includegraphics[width=6cm,height=7cm,angle=0,bb=0 0 1000 1000]{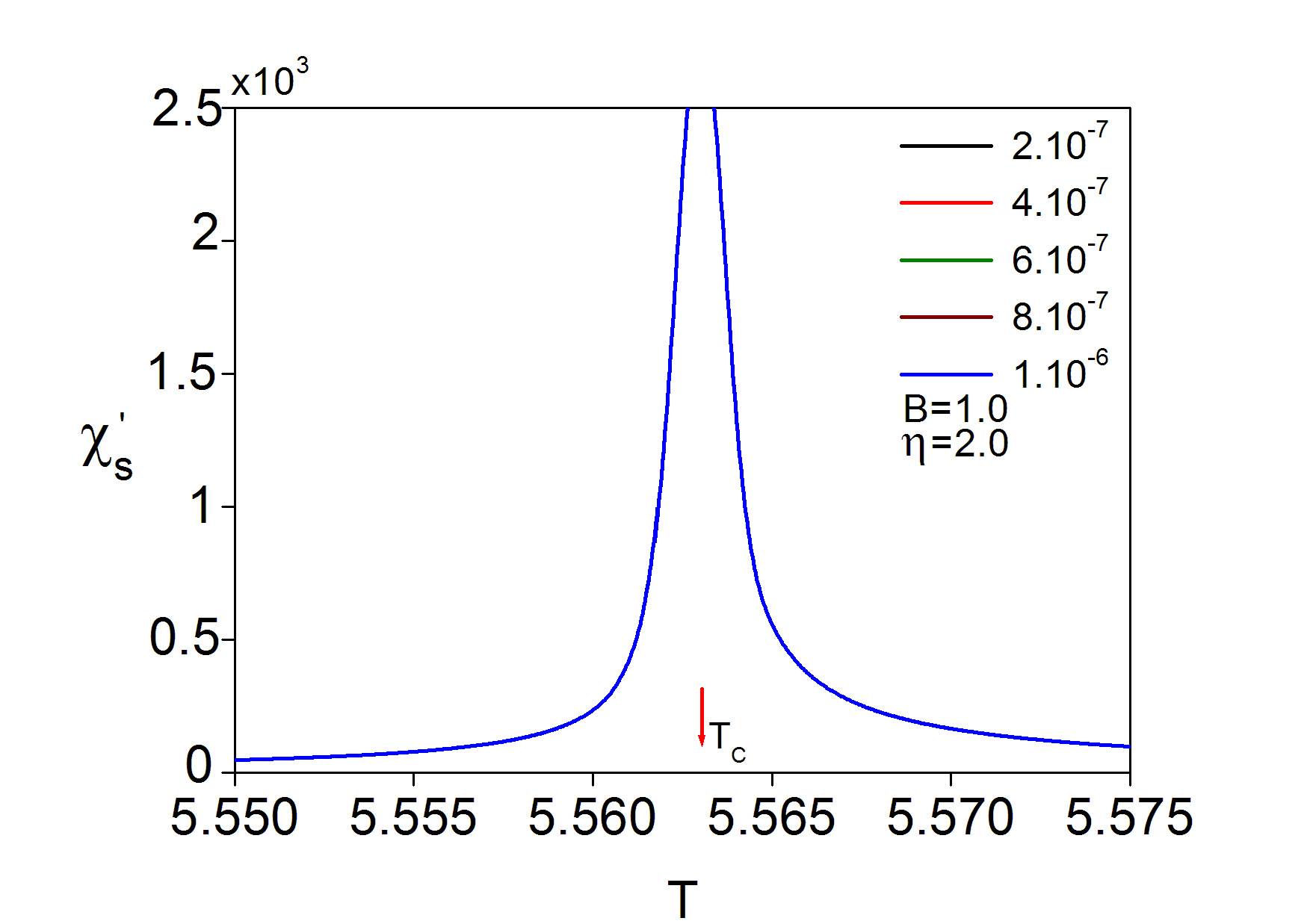}
  \includegraphics[width=6cm,height=7cm,angle=0,bb=0 0 1000 1000]{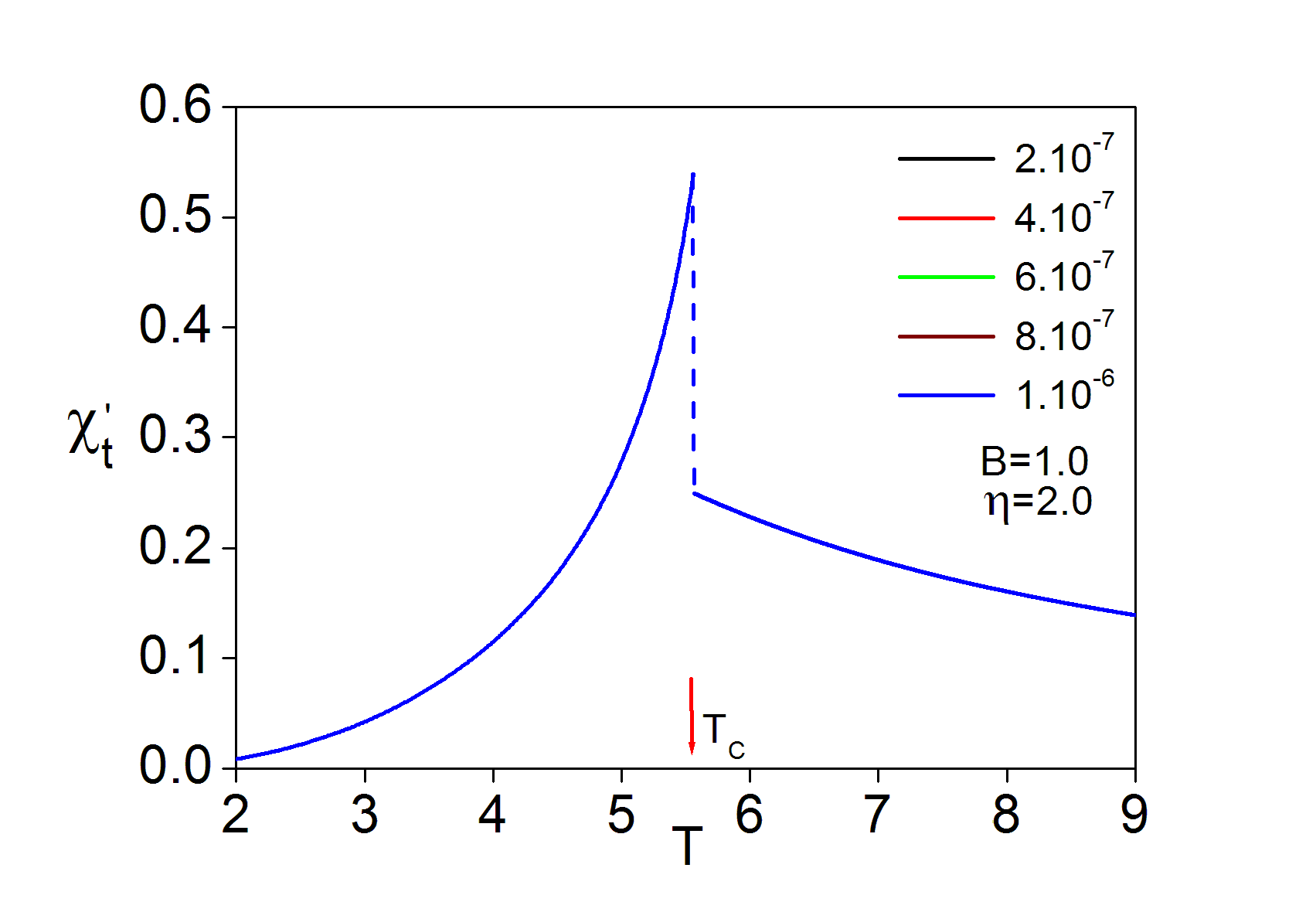}
 \caption{Temperature variations of staggered magnetic dispersion (a) and direct magnetic dispersion (b) for the low frequency region at several values of $\omega$ in the neighborhood second phase transition point.}
\end{center}
\end{figure}

\begin{figure}[tbp]
\begin{center}
 \includegraphics[width=6cm,height=7cm,angle=0,bb=0 0 1000 1000]{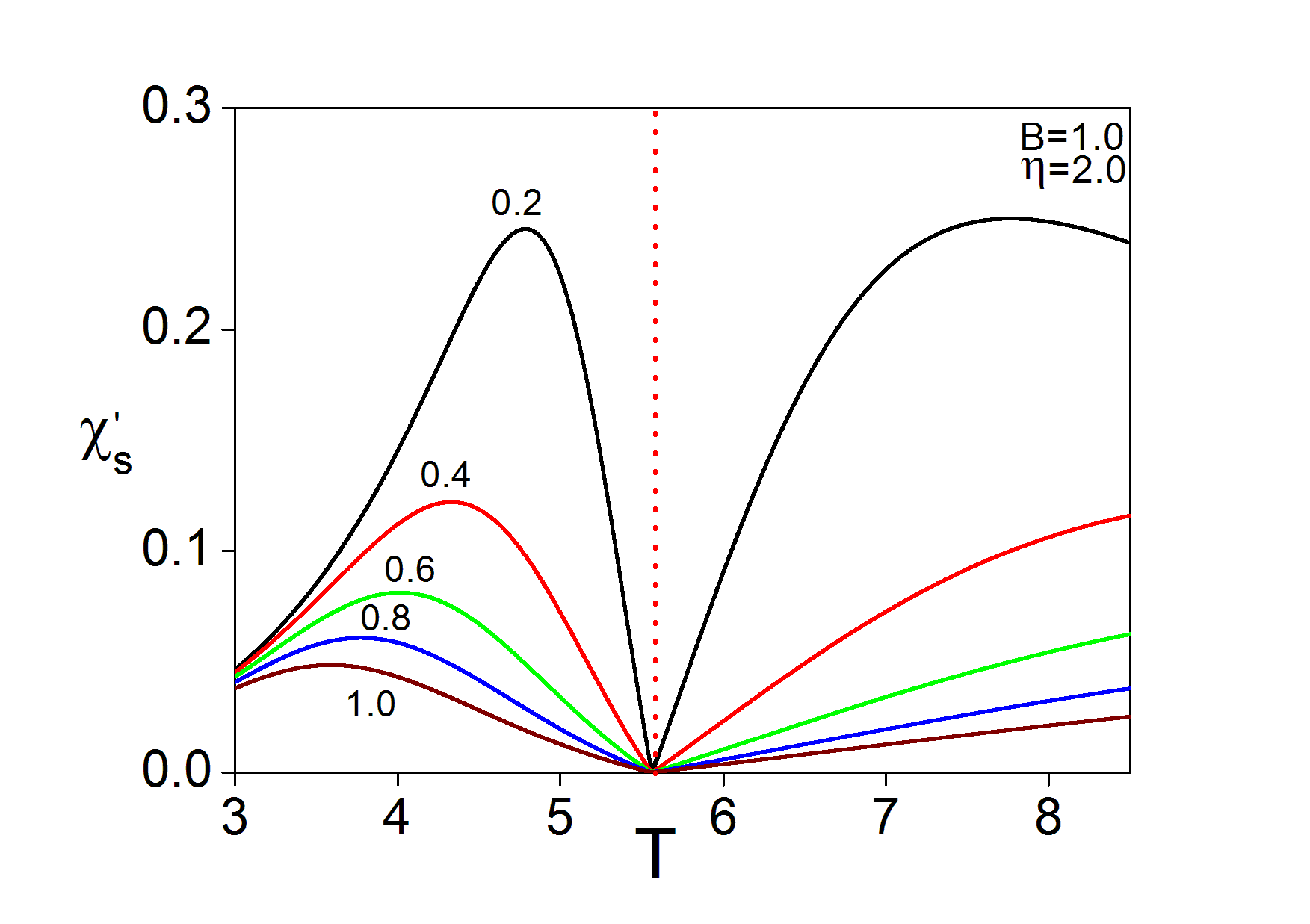}
  \includegraphics[width=6cm,height=7cm,angle=0,bb=0 0 1000 1000]{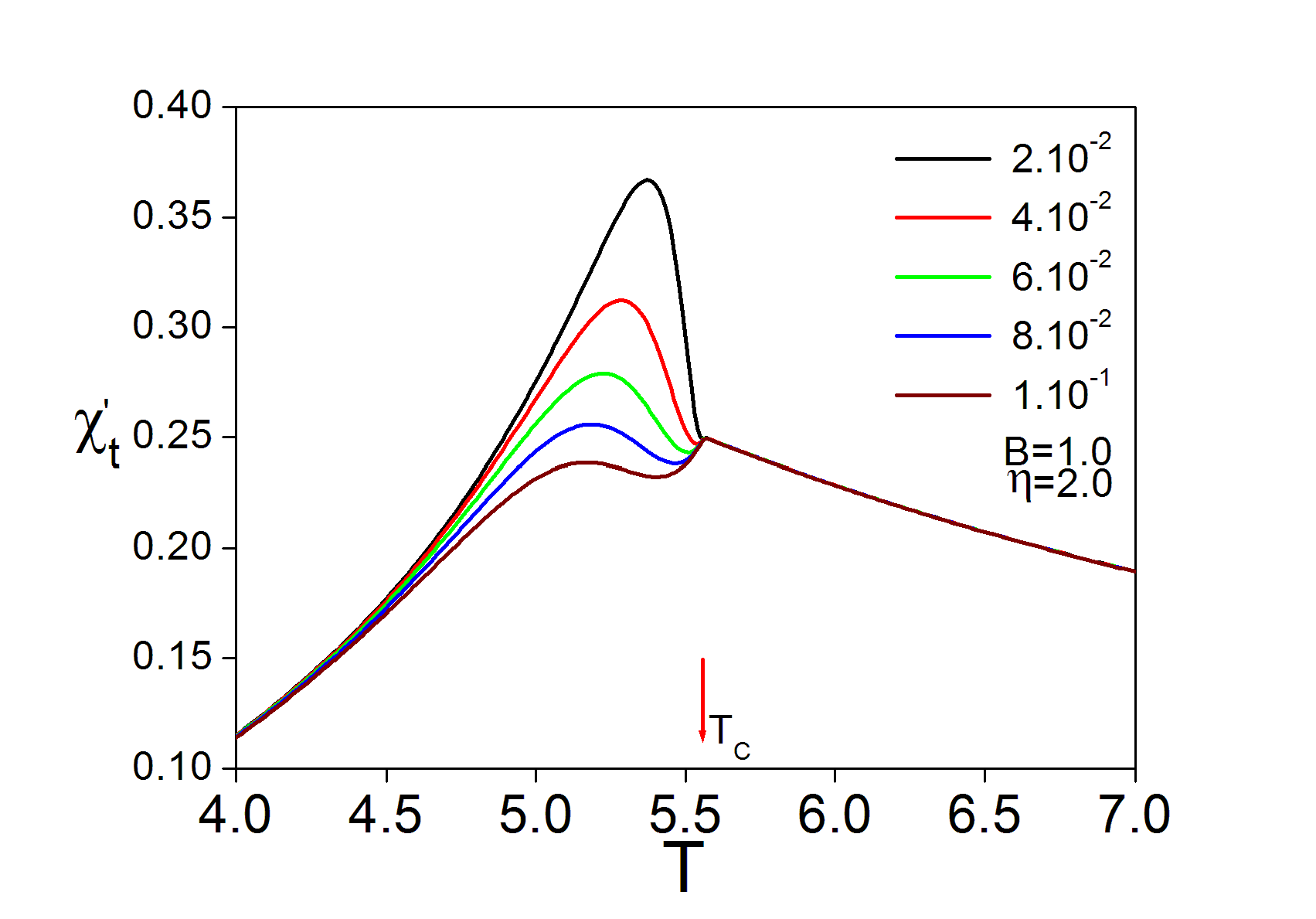}
 \caption{Temperature variations of staggered magnetic dispersion (a) and direct magnetic dispersion (b) for the high frequency region at several values of $\omega$ in the neighborhood second phase transition point. The number accompanying each curve denotes the frequency.}
\end{center}
\end{figure}

\begin{figure}[tbp]
\begin{center}
 \includegraphics[width=6cm,height=7cm,angle=0,bb=0 0 1000 1000]{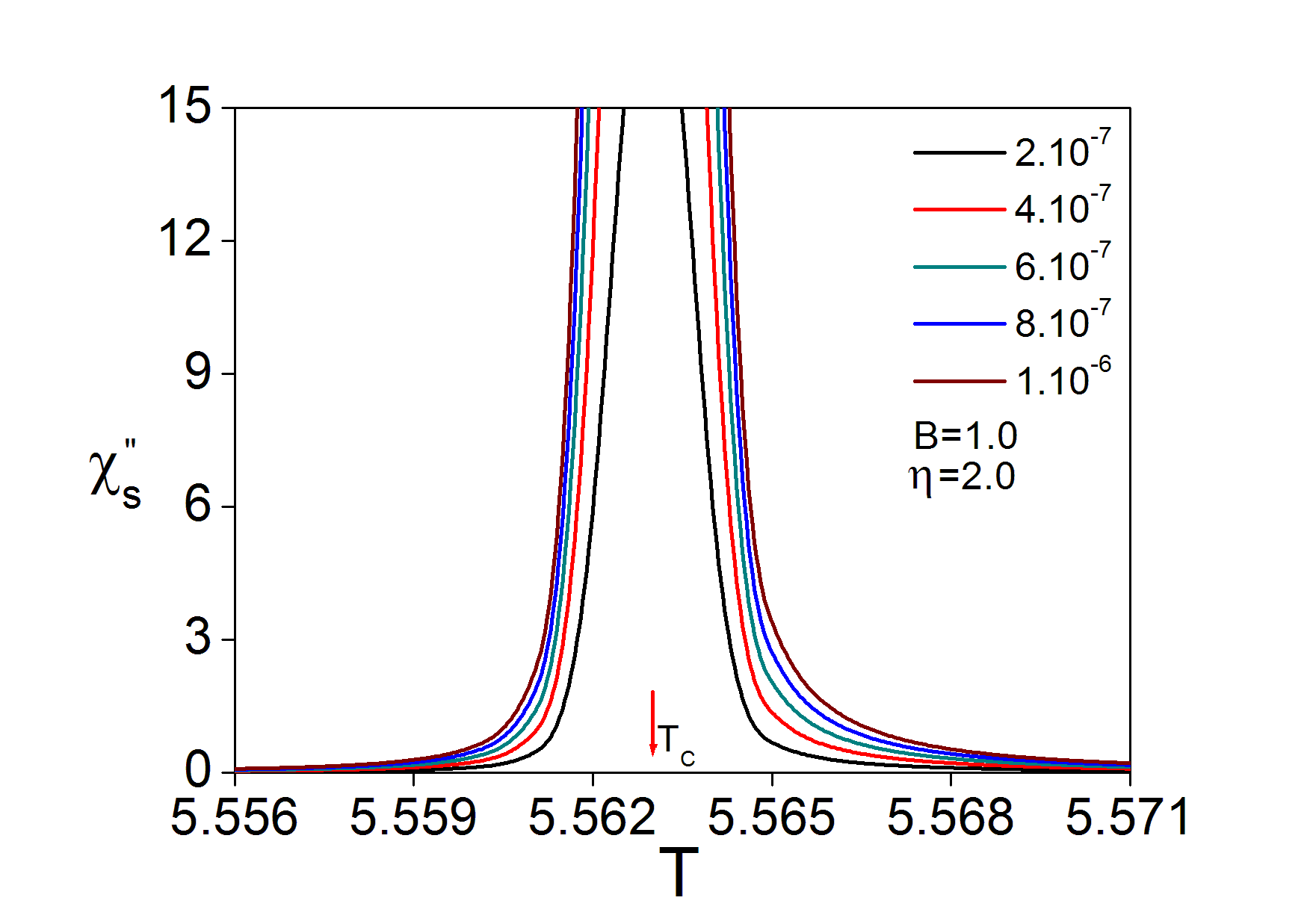}
  \includegraphics[width=6cm,height=7cm,angle=0,bb=0 0 1000 1000]{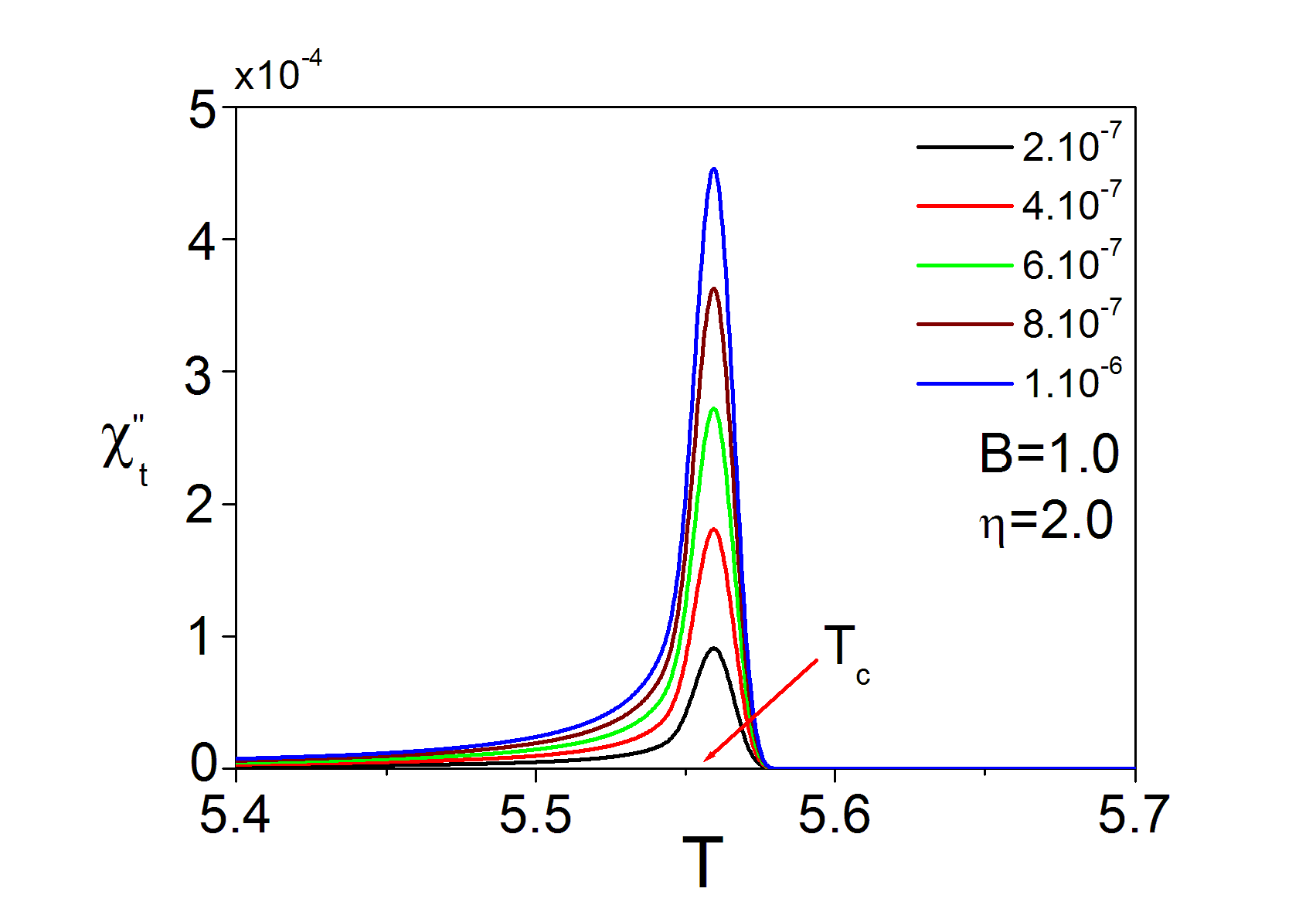}
 \caption{Temperature variations of staggered magnetic absorption (a) and direct magnetic absorption (b) for the low frequency region at several values of $\omega$ in the neighborhood second phase transition point.}
\end{center}
\end{figure}

\begin{figure}[tbp]
\begin{center}
 \includegraphics[width=6cm,height=7cm,angle=0,bb=0 0 1000 1000]{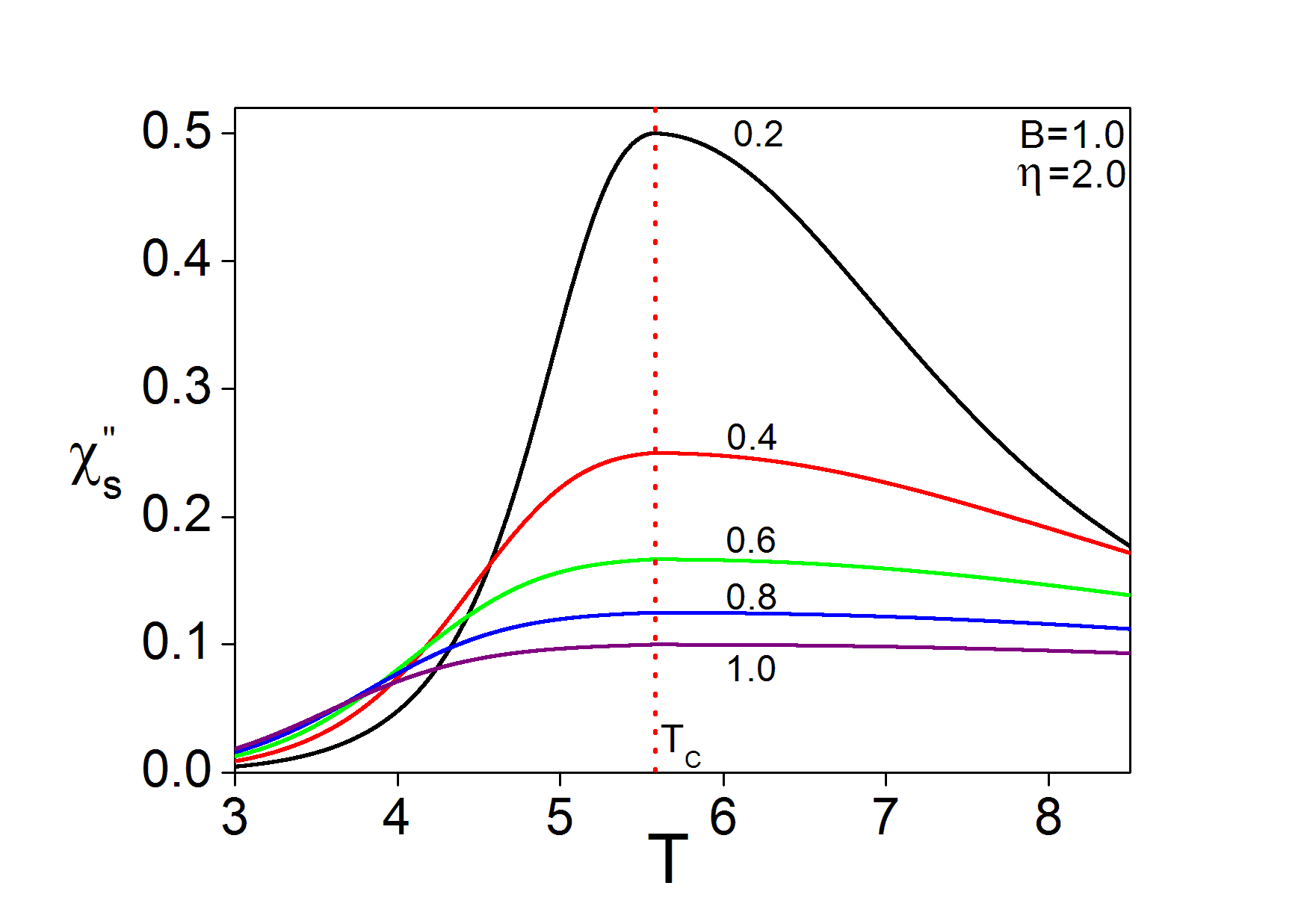}
  \includegraphics[width=6cm,height=7cm,angle=0,bb=0 0 1000 1000]{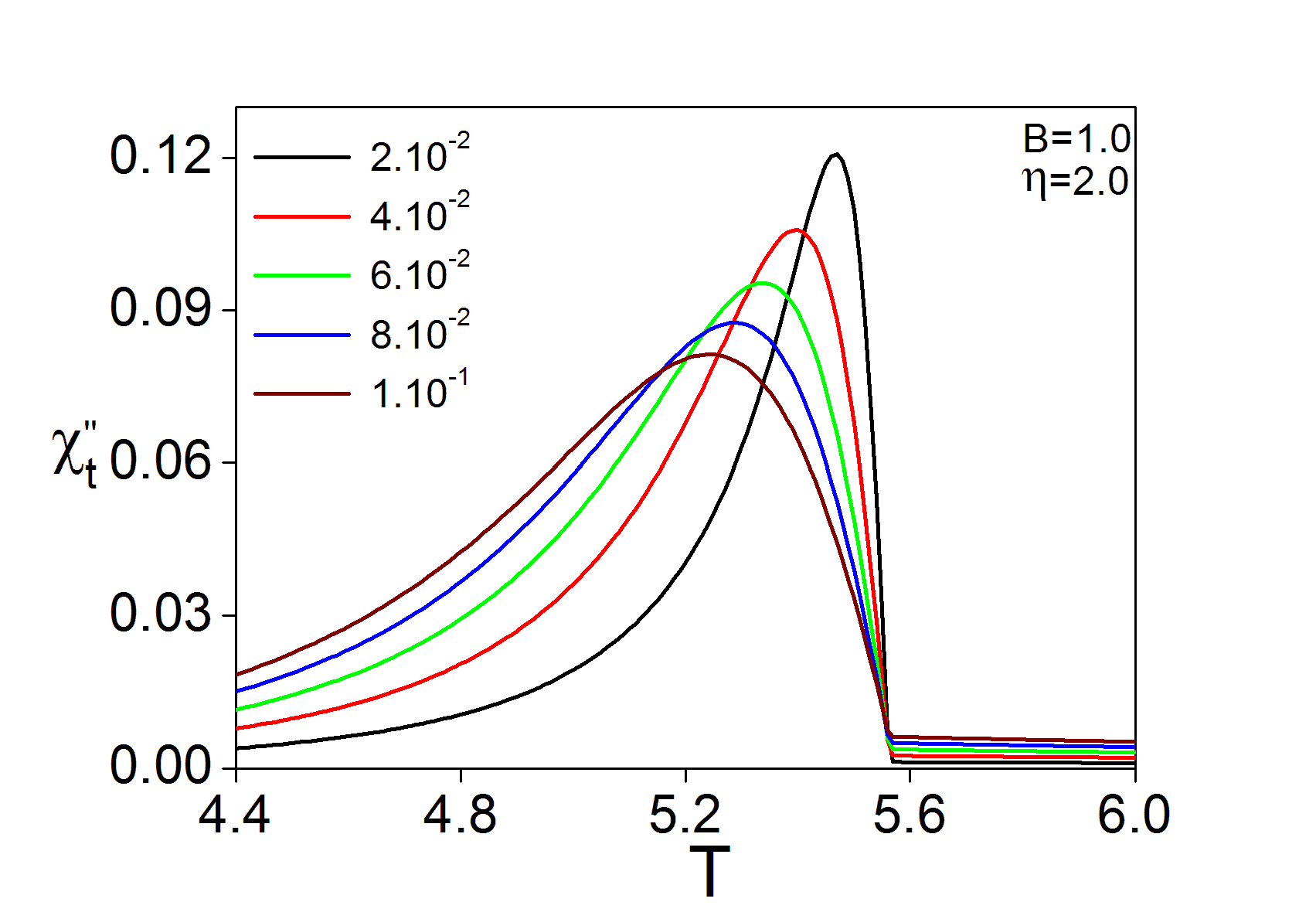}
 \caption{Temperature variations of staggered magnetic absorption (a) and direct magnetic absorption (b) for the high frequency region at several values of $\omega$ in the neighborhood second phase transition point. The number accompanying each curve denotes the frequency.}
\end{center}
\end{figure}

\begin{figure}[tbp]
\begin{center}
  \includegraphics[width=6cm,height=7cm,angle=0,bb=0 0 1000 1000]{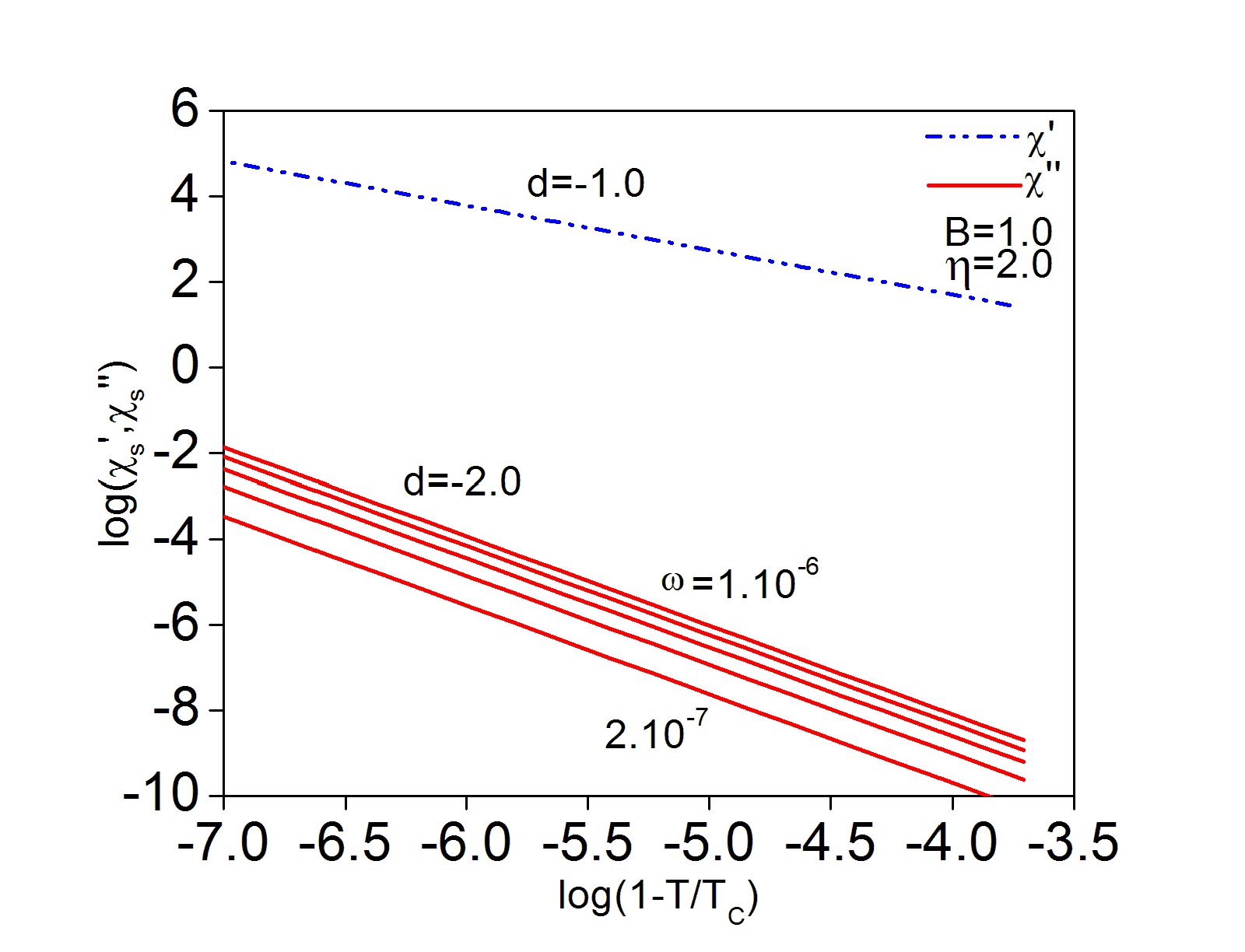}
  \includegraphics[width=6cm,height=7cm,angle=0,bb=0 0 1000 1000]{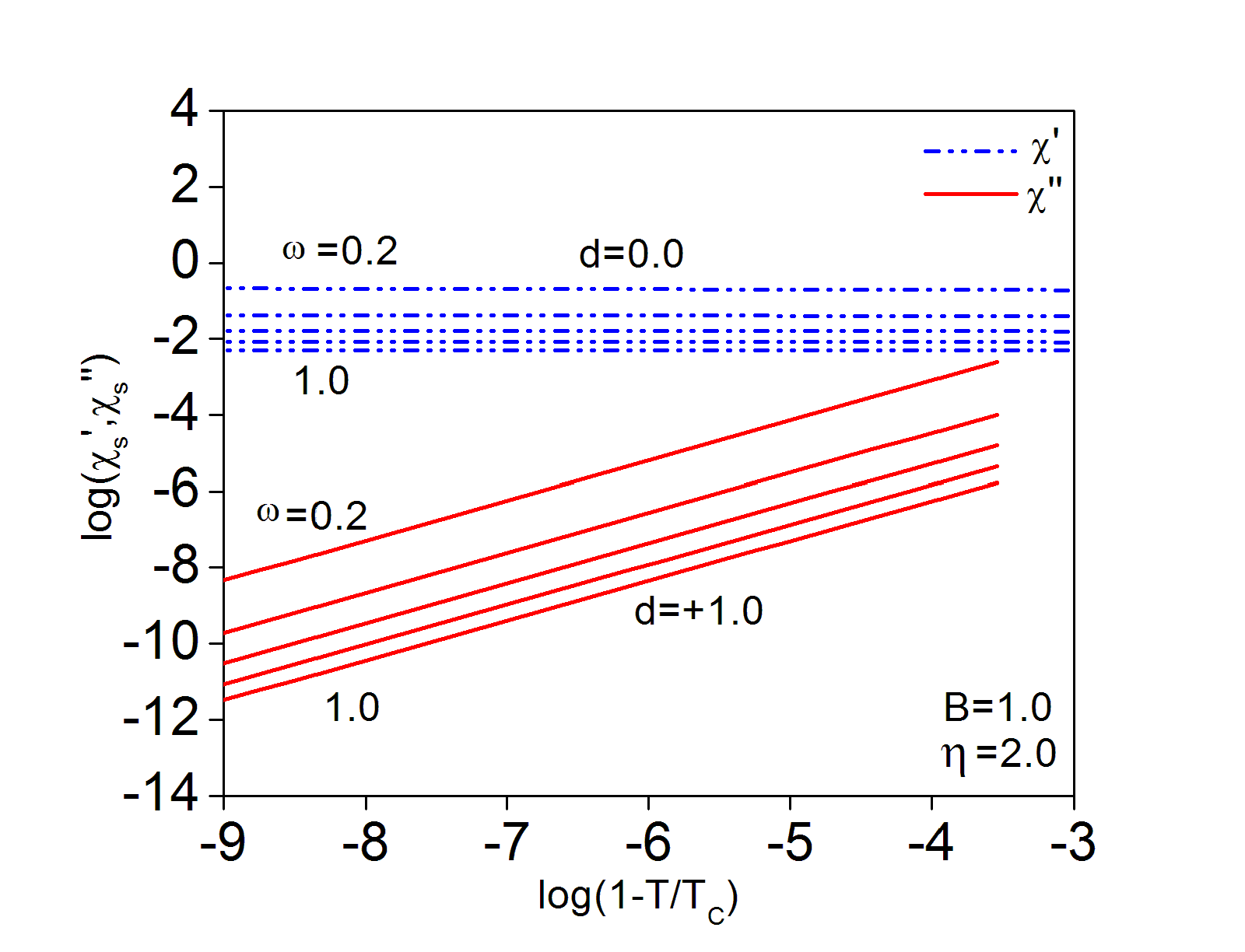}
  \includegraphics[width=6cm,height=7cm,angle=0,bb=0 0 1000 1000]{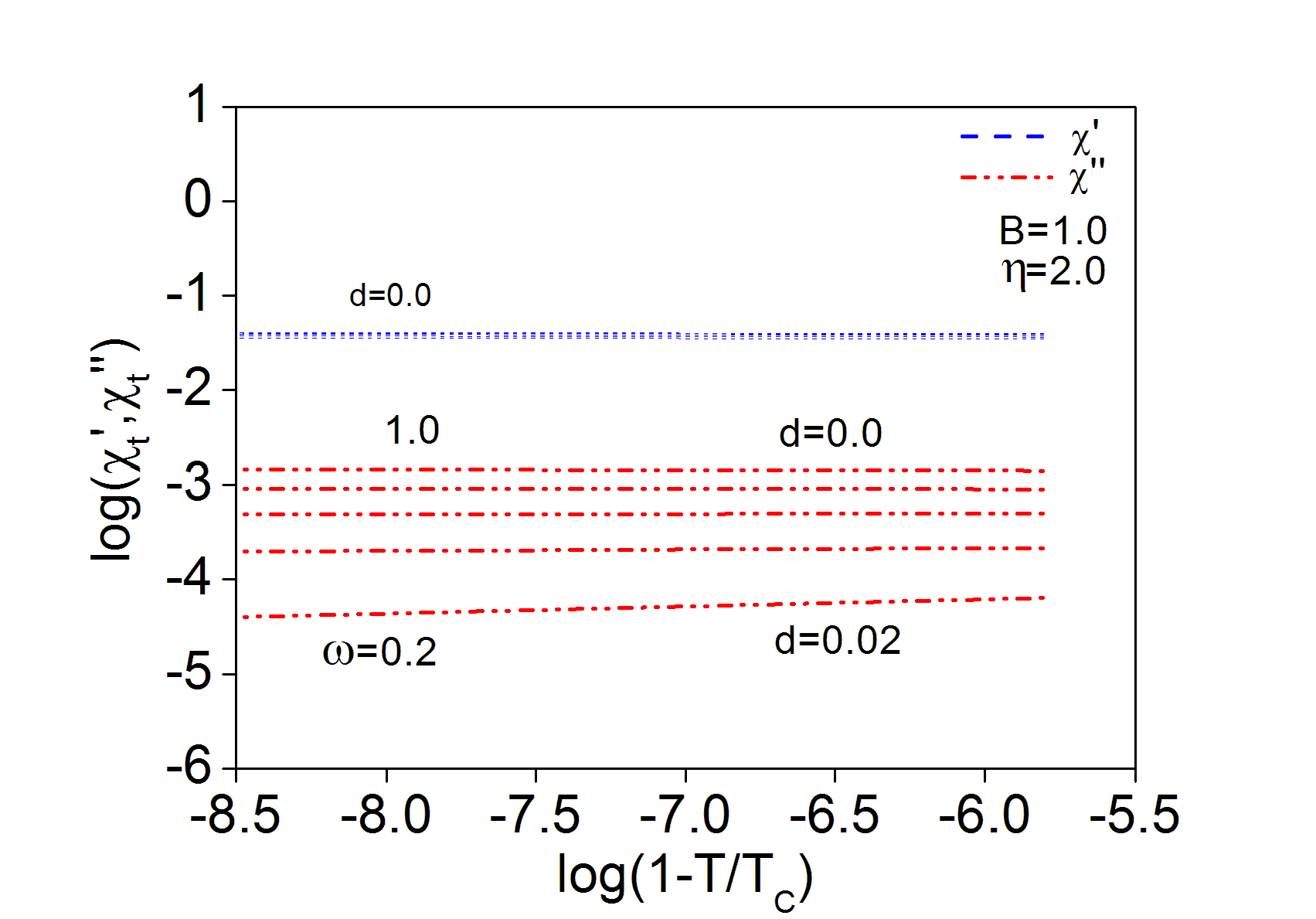}
 \caption{Logarithmic plot of staggered magnetic dispersion and staggered  magnetic absorption vs. reduced temperature $(1-T/T_{C})$ for $T<T_{C}$ at several values in the low frequency region (a) and high frequency region (b). Dotted lines are for $\chi_{s}^{'}$ and solid lines for $\chi_{s}^{''}$. }
\end{center}
\end{figure}

\clearpage
\begin{table}
\centering
\caption{ Mean field critical exponents for  staggered magnetic dispersion, staggered magnetic absorbtion, direct magnetic dispersion  and  direct magnetic absorbtion in the low and high frequency regions.}
\vspace{0.2cm}
\label{T1}
\begin{tabular}{|c|c|c|c|}
  \hline
  Quantity & Frequency Range & Exponent Value & Critical Singularity \\
  \hline
  $\chi_{s}^{'}(\omega)$ & $2.10^{-7}-1.10^{-6}$ (low freq.) & -1 & divergence \\
  \hline
  $\chi_{s}^{''}(\omega)$& $2.10^{-7}-1.10^{-6}$ (low freq.) & -2 & divergence \\
  \hline
  $\chi_{s}^{'}(\omega)$  & $0.2-1.0$ (high freq.) & 0 & cusp \\
  \hline
  $\chi_{s}^{''}(\omega)$ & $0.2-1.0$ (high freq.) & +1 & cusp \\
  \hline
  $\chi_{t}^{'}(\omega)$ & $0.2$ (high freq.) & 0.02 & cusp \\
  \hline
  $\chi_{t}^{'}(\omega)$ & $0.4-1.0$ (high freq.) & 0 & cusp \\
  \hline
  $\chi_{t}^{''}(\omega)$ & $0.2-1.0$ (high freq.) & 0 & cusp \\
  \hline
\end{tabular}
\end{table}


\begin{thebibliography}{99}
\bibitem {Engelstad} P. E. Engelstad and K. Yamada, Phys. Rev. B 52 (1995) 13029.
\bibitem {Durin} G. Durin, M. Bonaldi, M. Cerdonio, R. Tommasini, S. Vitale, J. Magn. Magn. Mater. 101 (1991) 89.
\bibitem {Raap} M.B.F. van Raap, F.H. Sanchez, C.E.R. Torres, L. Casas, A. Roig, E. Molins, J. Phys.: Condens. Matter 17 (2005) 6519.
\bibitem {Kötzler} J. Kötzler, G. Eiselt, J. Phys. C 12 (1979) L469.
\bibitem {Girtu}  M.A. Girtu, J. Optoelect. Adv. Mater. 4 (2002) 85.
\bibitem {Fannin}   P.C. Fannin, C.N. Marin, I. Malaescu, A.T. Giannitsis, J. Magn. Magn. Mater. 289 (2005) 78.
\bibitem {Barry66} J.H. Barry, J. Chem. Phys. 45 (1966) 4172.
\bibitem {Barry71} J.H. Barry, D.A. Harrington, Phys. Rev. B 4 (1971) 3068.
\bibitem {Suzuki} M. Suzuki and R. Kubo, J. Phys. Soc. Jpn. 24 (1968) 51.
\bibitem {Acharyya}  M. Acharyya, B.K. Chakrabarti, Phys. Rev. B 52 (1995) 6550.
\bibitem {Erdem} R.Erdem, J.Magn. Magn. Mater. 320 (2008) 2273.
\bibitem {Erdemfreq} R.Erdem, J.Magn. Magn. Mater. 321 (2009) 2592.
\bibitem {Stryjewski} E.Stryjewski, N. Giordano, Adv. Phys. 26 (1977) 487.
\bibitem{Landau}  D.P. Landau, Phys. Rev. Lett. 28 (1972) 449; B.L. Arora, D.P. Landau, AIP Conf. Proc. 10 (1973) 870.
\bibitem{Harbus} F. Harbus, H.E. Stanley, Phys. Rev. B 8 (1973) 1156; F. Harbus, H.E. Stanley, Phys. Rev. B 8 (1973) 1141.
\bibitem {Hernandez}  L. Hernandez, H.T. Diep, D. Bertrand, Phys. Rev. B 47 (1993) 2602.
\bibitem{Onyszkiewicz1} Z. Onyszkiewicz, A. Wierzbicki, Physica B 151 (1988) 462.
\bibitem{Onyszkiewicz2} Z. Onyszkiewicz, A. Wierzbicki, Physica B 151 (1988) 475.
\bibitem{das} D. Das, M. Barma, Physica A 270 (1999) 245.
\bibitem{sahni} P. S. Sahni, J. D. Gunton, S. L. Katz, R. H. Timpe, Phys. Rev. B 25 (1982) 389.
\bibitem{yan} Z. R. Yang, Phys. Rev. B 46 (1992) 11578.
\bibitem{oguz} E.Oguz, J. Phy. A 21 (1988) 2799 .
\bibitem{gulphysicA} G.Gulpinar, D.Demirhan, F. Buyukkilic, Physica A 383 (2007) 372.
\bibitem{gulPLAglauber} G.Gulpinar, D.Demirhan, F. Buyukkilic, Phys. Lett. A 373 (2009) 511.
\bibitem {GulPrevE} G. Gulpinar, D. Demirhan, F. Buyukkilic, Phys. Rev. E 75 (2007) 021104.
\bibitem {GulYenal} G. Gulpinar, Y. Karaaslan, Phys. Lett. A, 375 (2011) 978–983.
\bibitem {GulAtenuasyon} G. Gulpinar, Phys. Lett. A 372 (2008) 98.
\bibitem {Cohen} J. M. Kincaid and E. G. D. Cohen, Phys. Reports 22 (1975) 57.
\bibitem {Selke} W. Selke, Z. Phys. B 101 (1996) 145.
\bibitem {Zukovic1} M. Zukovic, A. Bobak and  T. Idogaki, J. Magn. Magn. Mater. 188 (1998) 52.
\bibitem {Zukovic2} M. Zukovic, A. Bobak and T. Idogaki, J. Magn. Magn. Mater. 192 (1999) 363.
\bibitem{Kasteleyn}  P.W. Kasteleyn,  Physica 22 (1956) 387 .

\bibitem {FisherandSykes} M.E. Fisher and M.F. Sykes, Physica 28 (1962) 919 ; 28, 939 (1962).
\bibitem{barry} J. H. Barry, D. A. Harrington, Phys. Rev. B 45 (1971) 3068.
\bibitem{groot} S.R. Groot and P. Mazur, Nonequilibrium Thermodynamics,
Amsterdam, North Holland Pub. Co. (1961).
\bibitem{barry_F}  J.H. Barry, J. Chem. Phys. 45 (1966) 4172.
\bibitem{Onsager} L. Onsager, Phys. Rev. 37, (1931) 405; {37} (1931) 2265.
\bibitem {Kikuchi} R. Kikuchi, Ann. Phys. 10 (1960) 127 ; (1963) Hughes Research Rep.
No. 271.

\end{thebibliography}
\end{document}